\documentclass[aps,prb,reprint,superscriptaddress,floatfix]{revtex4-2}
\usepackage{amsmath,amssymb,bm,mathtools}
\usepackage{microtype}
\usepackage{graphicx}
\usepackage[dvipsnames]{xcolor}
\usepackage{hyperref}
\hypersetup{
    colorlinks=true,
    citecolor= blue,
    linkcolor= blue,
    filecolor=blue,      
    urlcolor= blue,
    pdfcreator = {\LaTeX\ and \flqq hyperref\frqq},
}

\newcommand{\Ztwo}{\mathbb Z_2}
\newcommand{\ii}{\mathrm{i}}

\newcommand{\calS}{\mathcal S}

\newcommand{\sgn}{\operatorname{sgn}}

\begin{document}

\title{Exact selection of a toric-code vison crystal in a flux-conditioned Kitaev model}

\author{Jiucai Wang}
\affiliation{School of Physics, Hangzhou Normal University, Hangzhou 311121, China}

\author{Chuan Chen}
\email{chenc@lzu.edu.cn}
\affiliation{School of Physical Science and Technology, Lanzhou University, Lanzhou 730000, China}
\affiliation{Lanzhou Center for Theoretical Physics, Key Laboratory of Quantum Theory and Applications of MoE,
Gansu Provincial Research Center for Basic Disciplines of Quantum Physics, Lanzhou University, Lanzhou 730000, China}

\date{\today}

\begin{abstract}
We construct an exactly solvable extension of the spin-$1/2$ Kitaev honeycomb model in which conserved $\mathbb{Z}_2$ fluxes determine not only the signs but also the connectivity of nearest-neighbor Majorana hopping. At a tuned loop point, hopping survives only across opposite-flux plaquettes, so every vertex has active degree zero or two and the matter Hamiltonian fragments in each flux sector into independent Majorana rings and isolated zero modes. Exact ring spectra and bond counting then bound the matter energy over all local flux configurations and all four Wilson-loop sectors, and split it exactly into a frustrated triangular-lattice Ising term, whose extensively degenerate ground states are the fully packed loop coverings, and a non-negative Majorana residual. On admissible commensurate tori, the residual selects precisely the three translation-related $2/3$-vison crystals, which saturate the bound, and an exact fermion-parity identity shows that for antiferromagnetic coupling their vacua also survive projection, making them rigorous ground states of the spin model. Within a single crystal, a depth-one local unitary then maps the ground space onto that of a sheared square-lattice toric code, giving fourfold topological degeneracy. The model thus realizes exact nonperturbative gauge-matter feedback: the flux fixes where the Majorana fermions may move, and their zero-point energy selects in return a topologically ordered vison crystal with spontaneously broken translation symmetry.
\end{abstract}

\maketitle

\section{Introduction}
\label{sec:introduction}

The Kitaev honeycomb model provides an exactly solvable setting in which
emergent Majorana fermions propagate in a static $\Ztwo$ gauge
background~\cite{Kitaev2006}.  Within each gauge sector the matter Hamiltonian
is quadratic, but the gauge field modifies only the signs of the
nearest-neighbor hopping amplitudes: the hopping graph remains the honeycomb
lattice in every sector.  Matter energetics can nevertheless distinguish among
flux sectors, as exemplified by the flux-free ground state selected by Lieb's
theorem~\cite{Lieb1994}.  This raises a natural question: can conserved fluxes
control the connectivity of the Majorana network as well as the signs, and can
the ensuing gauge-matter feedback select a nonuniform flux ground state?

Conventional flux-dependent path interference suggests one route toward such
graph-level control: two spin strings along distinct paths with common
endpoints reduce to the same Majorana bilinear, with a relative sign fixed, up
to a geometrical factor, by the flux they enclose.  In solvable extensions of
the Kitaev model this makes third-neighbor hopping flux dependent and can
stabilize three-sublattice vison crystals, but leaves the nearest-neighbor
honeycomb network intact~\cite{ZhangPRL2019,ZhangPRR2020}.  Here we instead
dress the nearest-neighbor bond operator itself with the fluxes of its two
adjacent plaquettes.  The single-plaquette dressings are literal complementary
paths, whereas the two-plaquette dressing is not a spin string on any path.  
In a fixed flux sector all of them still contribute to the one
nearest-neighbor Majorana bilinear, with coefficients set by the two adjacent
fluxes, so at tuned couplings they cancel exactly on one or another flux
environment---deleting those bonds and thereby changing the kinetic graph.  
The resulting \emph{flux-conditioned Kitaev model} is exactly solvable despite
its two-, six-, and eight-spin interactions.

Its simplest limit is the loop point, where the cancellation is complete on
every bond joining equal fluxes: a bond is active only if its two adjacent
plaquettes carry opposite fluxes.  The three fluxes meeting at a vertex are
either all equal or split two to one, so every vertex has active degree zero or
two. The active graph therefore consists of disjoint closed loops and isolated
sites, reducing the matter Hamiltonian exactly to independent Majorana rings
and zero modes---a structure we term \emph{flux-conditioned graph fragmentation}.  The flux domain
walls fix the loop geometry, while each ring's $\Ztwo$ twist is set by the
fluxes it encloses, and for noncontractible rings by the global holonomy as
well, so identically shaped rings can exhibit different spectra.

This reduction makes the global flux competition analytically tractable.  Fully
packed loop configurations are precisely the ground states of the
triangular-lattice Ising antiferromagnet~\cite{Wannier1950,BloteNienhuis1994},
an extensive manifold that the global even-vison constraint restricts without
lowering its entropy density.  For a ring of any even length and $\Ztwo$ twist we obtain
the exact ground-state energy and a sharp lower bound per active bond, saturated
only by an antiperiodic elementary hexagon.  Combined with an exact
bond-counting identity, the bound splits the unprojected matter energy into a
triangular-Ising term that selects full packing and a nonnegative Majorana
residual that selects a gapped $2/3$-vison crystal of elementary rings, unique
up to three translations.  The degeneracy is thus lifted by fermionic zero-point
energy with no small parameter---a nonperturbative counterpart of
order-by-disorder~\cite{Villain1980,Shender1982,Chalker1992}.

Physical-state projection then decides whether this minimum is attained by the
spin model, and on finite tori it could exclude the unprojected matter ground
state.  We derive an exact fermion-parity identity and verify it against spin
exact diagonalization on six commensurate tori.  For antiferromagnetic coupling
the Majorana vacuum survives projection, so the crystal saturates the global
bound in the physical spin Hilbert space; on admissible tori the ground space is
twelvefold degenerate, three translation-related crystals, each with four Wilson
sectors.  For ferromagnetic coupling with an odd number of Majorana rings, projection
instead requires one matter excitation and can reorder finite-size levels.

Within a single crystal, a depth-one local unitary followed by edge relabeling
maps the conserved flux constraints onto the plaquette and star stabilizers of a
sheared square-lattice toric code~\cite{Kitaev2003}.  The equivalence is exact,
giving two logical qubits and hence fourfold topological degeneracy,
corresponding to the $\nu=0$ Abelian class of Kitaev's sixteenfold
way~\cite{Kitaev2006}.  The $2/3$-vison crystal therefore combines intrinsic
$\Ztwo$ topological order~\cite{Wen1990,XChen2010,Wen2017} with spontaneously broken translation
symmetry, rather than realizing a
translation-invariant quantum spin
liquid~\cite{Savary2017-bs,Zhou2017-pe,Knolle2019-hn,Broholm2020-pq,Balents2010-td}.

The loop network also organizes the defect physics: a local flux change can
destroy a ring and expose zero modes, or reconnect several rings without
producing any, so the zero modes track the induced connectivity rather than an
isolated vison, in contrast to the vison-bound Majorana zero modes of Ising topological order~\cite{Kitaev2006}.
The exact graph decomposition and the global selection bound
are specific to the fine-tuned loop point.
Away from it,
only sector-specific bounds remain, while finite-torus scans identify candidate regimes but do not establish thermodynamic phase boundaries.
Within these limits the model realizes a two-way organization of
gauge and matter: conserved flux determines the geometry of Majorana-fermion motion, and the
resulting matter energetics selects the flux pattern.

\section{Gauge-string construction}
\label{sec:construction}

\subsection{Kitaev model and flux conventions}
\label{subsec:kitaev_conventions}

\begin{figure}[t]
\centering
\includegraphics[width=\linewidth]{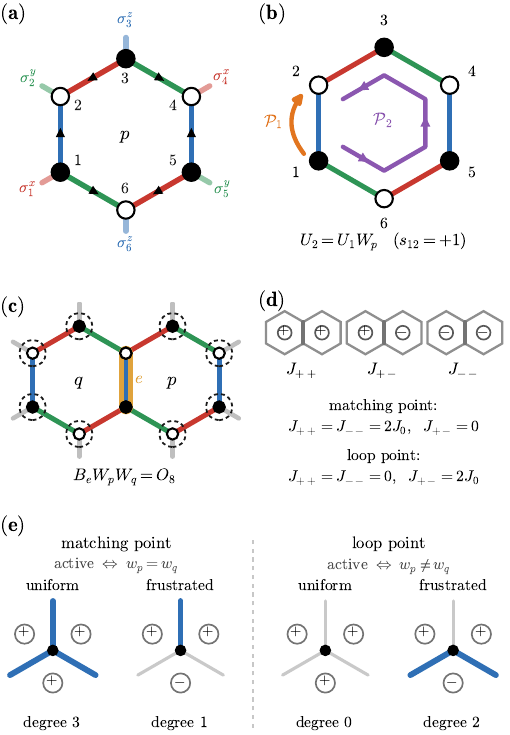}
\caption{(a) Plaquette and orientation conventions.  Filled (open) circles are
the $A$ ($B$) sublattice, arrows fix the $A\to B$ link orientation, red, green,
and blue mark $x$, $y$, and $z$ bonds, and outer labels give the Pauli factors
of $W_p$.
(b) Direct path $\mathcal{P}_1$ and complementary path $\mathcal{P}_2$ between
sites $1$ and $2$; their relative Wilson line is the enclosed flux $W_p$ up to
the geometrical sign $s_{12}$ of Eq.~\eqref{eq:pathratio}, here $+1$.
(c) Adjacent plaquettes $p$ and $q$ sharing bond $e$; dashed circles mark the
eight-spin support of $B_eW_pW_q$.
(d) The three local flux environments and the amplitudes assigned by the two
complementary flux-conditioned bond-cancellation limits, 
quoted as the signed coupling $2J_0$.
(e) The same rule on the dual triangle at a honeycomb vertex, for uniform and
frustrated flux configurations, with active bonds drawn thick.  The loop point
gives degrees $0$ and $2$, forcing nonbranching cycles and isolated vertices;
the matching point gives $3$ and $1$, hence branching clusters and dimers.
}\label{fig:gaugestring}
\end{figure}

We begin with the spin-$1/2$ Kitaev model on the honeycomb
lattice~\cite{Kitaev2006},
\begin{equation}
H_K=-\sum_{\langle ij\rangle_\alpha}K_\alpha B_{ij}^{\alpha},
\qquad
B_{ij}^{\alpha}=\sigma_i^\alpha\sigma_j^\alpha ,
\label{eq:HK}
\end{equation}
where $\alpha=x,y,z$ labels the three bond types.  As shown in
Fig.~\ref{fig:gaugestring}(a), the plaquette operator of each hexagon is
\begin{equation}
W_p=\sigma_1^x\sigma_2^y\sigma_3^z\sigma_4^x\sigma_5^y\sigma_6^z,
\label{eq:Wp_spin}
\end{equation}
Because $[B^\alpha_{ij},W_p]=0$, the plaquette operator $W_p$ is conserved under $H_K$.
Its eigenvalue $w_p=\pm1$ labels the local flux sector, with $w_p=-1$ a vison.

The spin operators can be represented in terms of Majorana fermions as
\begin{equation}
\sigma_i^\alpha=\ii b_i^\alpha c_i,
\qquad
D_i=b_i^x b_i^y b_i^z c_i=1.
\label{eq:majorana_rep}
\end{equation}
We orient every bond from $A$- to $B$-sublattice and define
\begin{equation}
u_{ij}=\ii b_i^\alpha b_j^\alpha,
\quad
u_{ji}=-u_{ij},
\quad
B_{ij}^{\alpha}=- \ii u_{ij} c_ic_j.
\label{eq:u_def}
\end{equation}
Therefore the plaquette flux takes the gauge-invariant form
\begin{equation}
W_p
=
\prod_{\substack{\langle ij\rangle\in\partial p\\
i\in A,\;j\in B}}
u_{ij}.
\label{eq:Wp_majorana}
\end{equation}
Individual link variables $u_{ij}$ are gauge dependent, whereas $W_p$ is not.
When a path traverses a bond opposite to the fixed $A\to B$
orientation, the corresponding link variable changes sign.  
Note that this orientation dependence must therefore be retained
when paths with common endpoints are compared.

\subsection{Open spin strings}
\label{subsec:open_strings}

Consider a simple path $\mathcal{P}: i_0\to i_1\to\cdots\to i_r$,
formed by $r$ consecutive Kitaev bonds, and define
$B_\ell\equiv B_{i_{\ell-1}i_\ell}$.
Adjacent bond operators anticommute, whereas nonadjacent ones commute; 
the prefactor $\ii^{\,r-1}$ below compensates the resulting phase and renders the
spin string Hermitian,
\begin{equation}
\mathcal{S}_{\mathcal{P}}
=
\ii^{\,r-1}B_1B_2\cdots B_r.
\label{eq:Spath}
\end{equation}
Evaluate each link variable in the direction of path traversal and define the
open Wilson line
\begin{equation}
U_{\mathcal{P}}=\prod_{\ell=1}^{r}u_{i_{\ell-1}i_\ell}^{(\mathcal{P})} .
\label{eq:openWilson}
\end{equation}
Substituting Eq.~\eqref{eq:u_def} then gives
\begin{equation}
\mathcal{S}_{\mathcal{P}}=-\ii U_{\mathcal{P}}c_{i_0}c_{i_r},
\label{eq:pathmap}
\end{equation}
since $\ii^{\,r-1}(-\ii)^{r}=-\ii$ independently of $r$ and the intermediate
matter Majoranas contract in pairs.  An extended spin string therefore reduces
exactly to an endpoint Majorana bilinear dressed by an open $\Ztwo$ gauge
string.  Although $U_{\mathcal{P}}$ is gauge dependent, the dressed bilinear
$U_{\mathcal{P}}c_{i_0}c_{i_r}$ is gauge invariant.

\subsection{Two paths and the enclosed Wilson loop}
\label{subsec:two_paths}

Consider two simple paths $\mathcal{P}_1$ and $\mathcal{P}_2$ with common
endpoints $i$ and $j$, such that the loop
$\mathcal{C}=\mathcal{P}_1\circ\overline{\mathcal{P}}_2$ is contractible, and
$R$ denotes the region it bounds.
The product of the plaquette Wilson loops inside $R$ cancels every interior link and yields
\begin{equation}
U_2=s_{12}U_1W_R,\quad
W_R\equiv\prod_{p\in R}W_p,\quad
s_{12}\equiv(-1)^{r_2+N_{\rm rev}(\mathcal C)}.
\label{eq:pathratio}
\end{equation}
Here $r_2$ is the length of $\mathcal{P}_2$ and
$N_{\rm rev}(\mathcal{C})$ counts contour links traversed opposite to
the fixed $A\to B$ orientation.  The geometrical sign $s_{12}$ is
gauge independent.

Substitution of Eq.~\eqref{eq:pathmap} shows that two spin strings with
common endpoints contribute to the same Majorana bilinear, with a
flux-dependent relative weight,
\begin{equation}
H_{ij}
=
-\ii U_1
\left(t_1+s_{12}t_2W_R\right)c_ic_j.
\label{eq:interference}
\end{equation}
where $t_a$ is the bare coupling that multiplies the spin-string
operator $\mathcal S_{\mathcal P_a}$ in the spin Hamiltonian ($a=1,2$).
Within a fixed flux sector, the effective amplitude is
\begin{equation}
t_{ij}^{\rm eff}
=
t_1+s_{12}t_2w_R.
\label{eq:teff}
\end{equation}
Consequently, the relative amplitude of the two paths is solely controlled by 
the enclosed $\mathbb Z_2$ flux.  For $|t_1|=|t_2|$, one flux sector
exhibits complete destructive interference.

For the elementary construction in Fig.~\ref{fig:gaugestring}(b), the two
paths are $\mathcal{P}_1:1\to2$, and $\mathcal{P}_2:1\to6\to5\to4\to3\to2$.
For $1,3,5\in A$ and $2,4,6\in B$, the orientation convention yields
\begin{equation}
U_2=U_1W_p,
\qquad
s_{12}=+1.
\end{equation}
Accordingly, a direct bond and its single-plaquette dressing contribute
$t_{ij}^{\rm eff}=t_1+t_2w_p$.

\section{Symmetric flux-conditioned Kitaev model}
\label{sec:model}

Motivated by the gauge-string construction, we now introduce
flux-dressed nearest-neighbor bond operators.  Specifically,
for a bond $e=\langle ij\rangle_\alpha$, $p_e$ and $q_e$ denote
the two adjacent plaquettes, and define
$B_e\equiv B_{ij}^{\alpha}=\sigma_i^\alpha\sigma_j^\alpha$.  
The minimal local interaction that treats the two adjacent plaquettes 
symmetrically~\footnote{The $p\leftrightarrow q$ symmetry is a genuine restriction.
Designating $p_e$ as the plaquette to the right of the $A\to B$ orientation admits 
the antisymmetric coupling $JB_e(W_{p_e}-W_{q_e})$, 
which keeps only flux-domain-wall bonds active but makes the hopping sign depend on
which side carries the vison.  We do not analyze this antisymmetric extension here.}
is
\begin{equation}
H
=
-\sum_e B_e
\left[
J_0
+J_1\left(W_{p_e}+W_{q_e}\right)
+J_2 W_{p_e}W_{q_e}
\right].
\label{eq:mainmodel}
\end{equation}
Here $B_e$ is the standard two-spin Kitaev interaction.  The products
$B_eW_{p_e}$ and $B_eW_{q_e}$ reduce to six-spin strings along the
complementary five-edge paths of the corresponding plaquettes and thus retain
the literal two-path interpretation of Sec.~\ref{subsec:two_paths}.  By
contrast, $B_eW_{p_e}W_{q_e}$ reduces to an eight-spin operator supported on the
outer boundary of the two-plaquette cluster 
[Fig.~\ref{fig:gaugestring}(c)]---a support that is disconnected, and therefore
not that of any spin string.  Its gauge factor is nonetheless the direct open
Wilson line times the gauge-invariant flux $W_{p_e}W_{q_e}$, so it contributes
to the same Majorana bilinear as $B_e$.  The mechanism below is accordingly
exact cancellation between flux-dressed bond operators, algebraically analogous
to conventional path interference but not requiring two simple paths.  With the
conventions of Eqs.~\eqref{eq:Wp_spin} and~\eqref{eq:Spath}, these identities
hold with the signs derived in Appendix~\ref{app:spinstrings}.

For uniform $J_0$, $J_1$, and $J_2$, Eq.~\eqref{eq:mainmodel} is invariant under time reversal and lattice translations. In the isotropic case, it also preserves the combined spin-lattice point-group symmetries of the Kitaev model. 
The $\mathbb Z_2$ link operators $u_{ij}$ remain conserved in the Majorana
representation, and each flux sector is again described by a quadratic
matter-Majorana Hamiltonian. 
Thus, in a fixed gauge sector, $W_p$ may be replaced by its eigenvalue $w_p=\pm1$, and the Hamiltonian is exactly reduced to
\begin{equation}
H_{\{u\}} = \ii\sum_{e=\langle ij\rangle} u_e\,J_e(\{w\})\,c_i c_j, \label{eq:majmodel}
\end{equation}
with the flux-conditioned nearest-neighbor amplitude
\begin{equation}
J_e(\{w\}) = J_0 +J_1\left(w_{p_e}+w_{q_e}\right) +J_2 w_{p_e}w_{q_e}. \label{eq:Jeff}
\end{equation}
Because a bond is adjacent to only two plaquettes, there are three
inequivalent local flux environments,
\begin{equation}
\begin{aligned}
J_{++}&=J_0+2J_1+J_2,\\
J_{+-}&=J_0-J_2,\\
J_{--}&=J_0-2J_1+J_2.
\end{aligned}
\label{eq:Jmm}
\end{equation}
Since $J_e$ depends only on the unordered pair $\{w_{p_e},w_{q_e}\}$, 
the $-+$ environment carries the same amplitude, $J_{-+}=J_{+-}$.
A $+-$ bond separates plaquettes carrying opposite fluxes and hence lies on a domain
wall of the dual Ising variables $w_p$, as shown in Fig.~\ref{fig:gaugestring}(d).

\subsection{Complementary limit: the matching point}
\label{subsec:matching}

Before turning to the central loop point, consider the complementary cancellation limit $J_0=J_2$. 
In this manifold, the $J_1$ term vanishes identically on every flux-domain-wall
bond, while the direct and two-plaquette-dressed terms cancel by construction.  Hence $J_{+-}=0$: all
Majorana hopping across flux domain walls vanishes exactly.
We call this the \emph{domain-wall-deletion manifold}: the spin lattice is unchanged, while the quadratic Majorana graph loses every domain-wall edge.

Setting $J_1=0$ further gives the \emph{matching point},
\begin{equation}
J_{++}=J_{--}=2J_0,\quad
J_{+-}=0,\quad
t_{\rm M}\equiv2|J_0|.
\label{eq:matchingpoint}
\end{equation}
Every bond between equal fluxes is then active with the same magnitude.
Since each honeycomb vertex corresponds to a dual triangle, its active degree
equals the number of equal-sign flux pairs: three when all three fluxes are
equal and one when they split two to one.  The active graph therefore consists
of branched components and isolated dimers, in contrast to the degree-zero-or-two
loop decomposition at the loop point [Fig.~\ref{fig:gaugestring}(e)].

On the triangular-Ising ground-state manifold, each triangle has exactly one
equal-sign pair, so every vertex has active degree one and the active graph is
a \emph{perfect matching} [Fig.~\ref{fig:gaugestring}(e)]---the dimer covering
of Sec.~\ref{sec:loops}, complementary to the fully packed
loops~\cite{MoessnerSondhi2001}.  Each Majorana dimer
$\ii t_{\rm M}u_ec_ic_j$ has unprojected ground-state energy $-t_{\rm M}$,
so every matching sector has energy density $-t_{\rm M}/2$ per site, and the
exponentially many matching sectors are exactly degenerate.

The degeneracy of the perfect-matching manifold does not by itself
 determine the global flux minimum. Instead, a global bound follows from 
the special form of the hopping matrix.  Let $A_{\rm K}[v]$ denote the
full-honeycomb Majorana matrix with uniform bond magnitude and
link signs $v_e$.  At the matching point,
\begin{equation}
A_{\rm match}[u,w]
=A_{\rm K}[u]+A_{\rm K}[u'],
\quad u'_e=u_e w_{p_e}w_{q_e}.
\label{eq:matching-decomposition}
\end{equation}
Both terms are uniform-magnitude hopping matrices on the full honeycomb graph, but
generally represent different flux sectors.

For $H=\frac{\ii}{4}\bm c^TA\bm c$, the unprojected vacuum energy is
$E_{\rm M}^{(0)}[A]=-\|\ii A\|_1/4$.  The trace-norm triangle inequality,
together with Lieb's theorem applied separately to the two full-graph
matrices, therefore gives
\begin{align}
E_{\rm M}^{(0)}[A_{\rm match}]
&\ge-\frac14\left(
\|\ii A_{\rm K}[u]\|_1+\|\ii A_{\rm K}[u']\|_1\right)\nonumber\\
&\ge-\frac12\|\ii A_{\rm zf}\|_1,
\label{eq:matching-global-bound}
\end{align}
where $A_{\rm zf}$ is the optimal zero-flux full-honeycomb matrix,
including the minimizing global holonomy.

For uniform zero flux with the optimal global holonomy, $u'=u$; hence the
trace-norm triangle inequality is saturated and both terms attain Lieb's bound.  
The flux-free sector therefore saturates
Eq.~\eqref{eq:matching-global-bound} and is a global minimizer of the
unprojected matter energy, rather than merely the lowest sampled sector.  
Its thermodynamic energy density,
$E_{\rm M}^{(0)}/N_s\simeq-0.7873t_{\rm M}$, lies below the perfect-matching
value $-t_{\rm M}/2$.  Whether any nonuniform flux configuration can satisfy
both equality conditions is not established here.

\subsection{The loop point}
\label{subsec:looppoint}

The central limit realizes the complementary connectivity: hopping survives
\emph{only} on flux-domain-wall bonds [see Fig.~\ref{fig:gaugestring}(d)].  It occurs at the \emph{loop point}
\begin{equation}
J_1=0,
\qquad
J_2=-J_0,
\label{eq:looppoint}
\end{equation}
for which $J_{++}=J_{--}=0$ and $J_{+-}=2J_0$. 
Upon defining $t_{\rm DW}\equiv2|J_0|$,
the spin Hamiltonian becomes
\begin{equation}
H_{\rm DW}^{\rm spin}
=
-J_0\sum_e B_e\left(1-W_{p_e}W_{q_e}\right).
\label{eq:HDWspin}
\end{equation}
Within a fixed gauge sector, the corresponding matter Hamiltonian is
\begin{equation}
H_{\rm DW}^{\rm M}
=
\ii t_{\rm DW}
\sum_{e:\,w_{p_e}\neq w_{q_e}}
\widetilde u_e\,c_i c_j,
\quad
\widetilde u_e=\operatorname{sgn}(J_0)u_e.
\label{eq:HDW}
\end{equation}
Hence, the active Majorana bonds coincide exactly with
the domain walls of the conserved plaquette fluxes.

\section{Exact Majorana-loop reduction}
\label{sec:loops}

The centers of the honeycomb plaquettes form a triangular lattice, on
which the conserved fluxes $w_p=\pm1$ may be viewed as Ising variables.
At the loop point defined by Eq.~\eqref{eq:looppoint}, a honeycomb bond is active
if and only if its two adjacent plaquettes carry opposite fluxes. 
Accordingly, the active Majorana edge set coincides with the domain-wall graph
of the dual Ising configuration.

Three distinct plaquettes meet at every honeycomb vertex and form one
elementary triangle of the dual lattice. 
Each honeycomb vertex corresponds uniquely to an elementary triangle of the dual triangular lattice.
The three Ising variables of a triangle are either all equal or split
two to one, so its number of unequal pairs is respectively zero or two.  Hence every
honeycomb vertex has active degree zero or two, and the active graph
consists of disjoint closed loops and isolated vertices [Fig.~\ref{fig:gaugestring}(e)].

With the matter Hamiltonian written as
\begin{equation}
H_{\rm DW}^{\rm M}
=
\frac{\ii}{4}\bm c^{T}A\,\bm c, \label{MajH}
\end{equation}
a permutation $P$ that orders the Majorana sites loop by loop gives
\begin{equation}
PAP^{T}
=
\bigoplus_\ell A_\ell
\oplus
0_{N_0\times N_0},
\label{eq:block_decomposition}
\end{equation}
where $A_\ell$ describes a loop of length $L_\ell$ and $N_0$ is the
number of isolated Majoranas. Since every active vertex has degree two,
the number of active bonds equals the number of active vertices
$N_{\rm act}=\sum_\ell L_\ell=N_s-N_0$.
We consider periodic clusters with distinct incident plaquettes and a globally
well-defined bipartite $A$--$B$ labeling, so all $L_\ell$ and $N_0$ are even.
The honeycomb lattice has girth six, so its shortest contractible cycle is an
elementary face.  We call a torus \emph{admissible} when it is commensurate
with the three-sublattice order and its shortest noncontractible cycle is
longer than six; the minimization and uniqueness statements below are
understood under this hypothesis.
Admissibility need not hold even for large systems: a torus that is extended in one direction but narrow in the other may remain inadmissible. 
For such tori, the uniqueness result of Sec.~\ref{sec:groundstate} is replaced by the direct sector enumeration described 
in Appendix~\ref{app:finite-torus-ed}.
The plaquette fluxes obey the global torus constraint $\prod_p w_p=+1$, implying an even total number of visons.

Each loop is characterized by its length $L_\ell$ and a gauge-invariant
twist $\varphi_\ell=\pm1$, equivalently the boundary condition
$c_{L_\ell+1}=\varphi_\ell c_1$ used in
Sec.~\ref{sec:ringspectrum}. Along a closed honeycomb loop, exactly
$L_\ell/2$ bonds are traversed opposite to the fixed $A\to B$
orientation, and thus
\begin{equation}
\varphi_\ell
=
(-1)^{L_\ell/2}
\prod_{e\in\ell}^{A\to B}u_e.
\label{eq:loop_twist_general}
\end{equation}
The expression is gauge invariant because every site contributes its
local gauge factor twice and it is unchanged when the loop orientation is reversed. 
For a contractible loop enclosing a region $R_\ell$, the twist reduces to
\begin{equation}
\varphi_\ell
=
(-1)^{L_\ell/2}
\prod_{p\in R_\ell}w_p .
\label{eq:loop_twist_flux}
\end{equation}
The global constraint ensures
that the result is unchanged if the complementary region is used instead.  

For a noncontractible loop, the twist is not determined by the local plaquette fluxes alone. 
Choosing a reference cycle $\gamma$ in the same homology class, define the Wilson loop, or global $\mathbb Z_2$ holonomy,
\begin{equation}\label{eq:global_holonomy}
\mathcal W_\gamma
\equiv \prod_{e\in\gamma}^{A\to B}u_e = \pm1.
\end{equation}
The strip between $\ell$ and $\gamma$ is denoted by $S_{\ell\gamma}$, then
\begin{equation}\label{eq:noncontractible_twist}
\varphi_\ell = (-1)^{L_\ell/2} \mathcal W_\gamma
\prod_{p\in S_{\ell\gamma}}w_p.
\end{equation}
Thus noncontractible loop spectra depend on both the local flux pattern
and the global $\mathbb Z_2$ holonomy.

Denoting by $E_0(L_\ell,\varphi)$ the ground-state energy of a
Majorana loop, as evaluated in Sec.~\ref{sec:ringspectrum},
the matter energy at the loop point takes the form
\begin{equation}
E_{\rm M}^{(0)}
\bigl[\{w_p\};\mathcal W_1,\mathcal W_2\bigr]
=
\sum_\ell
E_0(L_\ell,\varphi_\ell).
\label{eq:flux_energy_functional}
\end{equation}
Isolated Majoranas are exact zero modes and contribute no energy to
Eq.~\eqref{eq:flux_energy_functional}. The induced flux functional is
intrinsically nonlocal: $L_\ell$ depends on the extended geometry of a
domain wall, while $\varphi_\ell$ depends on the Wilson flux associated
with the entire loop.

When $N_0=0$, the fluxes around every elementary dual triangle are
$(+,+,-)$, up to an overall sign and permutation.  Each triangle therefore
contains one frustrated equal-sign bond and two satisfied opposite-sign bonds
of the triangular-lattice Ising antiferromagnet~\cite{Wannier1950}.  The
honeycomb edges crossing the frustrated bonds form a perfect matching, whereas
their complementary active edges form a fully packed Majorana-loop
covering~\cite{BloteNienhuis1994,Kondev1996}.

On a torus the allowed sectors are further restricted to the even-vison subset
of the extensively degenerate triangular-Ising ground-state manifold by
$\prod_p w_p=+1$, a single global constraint that leaves the entropy density
unchanged. 
Since $[H_{\rm DW}^{\rm spin},W_p]=0$, the flux pattern and loop
geometry are static in each sector.

\section{Exact Majorana-loop spectrum and induced flux energetics}
\label{sec:ringspectrum}

\subsection{Majorana-ring spectrum}
 
Consider one of the loops identified in Sec.~\ref{sec:loops}, with
Hamiltonian~\cite{Kitaev2001}
\begin{equation}
H_\ell
=
\ii t_{\rm DW}
\sum_{n=1}^{L}
s_n c_nc_{n+1},
\qquad
c_{L+1}=c_1,
\label{eq:ringH}
\end{equation}
where $L$ is even and $s_n=\pm1$. A local sign transformation
$c_n\to\eta_nc_n$ removes the individual $s_n$, leaving only their
gauge-invariant product
\begin{equation}
\varphi_\ell
=
\prod_{n=1}^{L}s_n
=
\pm1 .
\end{equation}
The transformation is orthogonal, and hence leaves the single-particle
spectrum unchanged.

\begin{figure}[t]
\centering
\includegraphics[width = \linewidth]{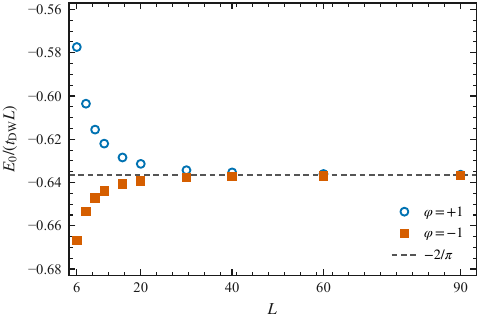}
\caption{The ground-state energy per site of an isolated Majorana loop. The two twist sectors approach the infinite-chain limit from opposite sides. Among loops with $L\ge6$, the $L=6$, $\varphi=-1$ ring has the lowest energy per site.
}\label{fig:loopenergy}
\end{figure}

After removal of the local signs, the twist enters through
$c_{L+1}=\varphi_\ell c_1$ and imposes the momentum quantization
\begin{equation}
k_m=
\begin{cases}
2\pi m/L, & \varphi_\ell=+1,\\[1mm]
(2m+1)\pi/L, & \varphi_\ell=-1.
\end{cases}
\label{eq:kquant}
\end{equation}
For Eq.~\eqref{MajH}, the eigenvalues of $\ii A$ are
\begin{equation}
\lambda(k)=-4t_{\rm DW}\sin k.
\end{equation}
Since $A$ is real and antisymmetric, the spectrum of $\ii A$ occurs in
pairs $\pm\epsilon_a$ (corresponding single-particle energies), here realized as $\lambda(-k)=-\lambda(k)$; each pair corresponds to one complex-fermion mode, and the unprojected Majorana vacuum has energy
$E_0=-\frac12\sum_{\epsilon_a>0}\epsilon_a$.  
It follows that the loop ground-state energies are (see Appendix~\ref{app:ring})
\begin{align}
E_0(L,+)
=
-2t_{\rm DW}\cot\frac{\pi}{L}, \\
E_0(L,-)
=
-2t_{\rm DW}\csc\frac{\pi}{L}.
\label{eq:ringenergy}
\end{align}
At fixed $L$, the $\varphi=-1$ sector is lower by
\begin{equation}
E_0(L,-)-E_0(L,+)
=
-2t_{\rm DW}\tan\frac{\pi}{2L}.
\end{equation}
As shown in Fig.~\ref{fig:loopenergy}, both branches approach $-2t_{\rm DW}/\pi$ per site as
$L\to\infty$, but from opposite sides. 
The $\varphi=-1$ sector is energetically favored for all finite $L$, with the splitting between the two sectors being most pronounced 
for short loops and vanishing in the large-$L$ limit.

Specializing Eq.~\eqref{eq:loop_twist_flux} to an elementary hexagon
gives $\varphi_{\rm hex}=-w_p$; the sign relation is verified explicitly in
Appendix~\ref{app:hexsign}. Hence, a flux-free hexagon has the energetically
favorable antiperiodic twist. For $L=6$,
\begin{equation}
E_0(6,-)=-4t_{\rm DW},
\quad
E_0(6,+)=-2\sqrt3\,t_{\rm DW}.
\label{eq:hexenergies}
\end{equation}
The $\varphi=+1$ hexagon has two zero eigenvalues of $\ii A$,
whereas the $\varphi=-1$ hexagon is gapped, with positive single-particle
energies $2t_{\rm DW}$, $2t_{\rm DW}$, and $4t_{\rm DW}$.

\subsection{Sharp energy bound and exact Ising decomposition}
\label{subsec:ising_decomposition}

For a fixed loop length, the antiperiodic sector has the lower energy,
$E_0(L,-)<E_0(L,+)$.  Together with $\sin(\pi/L)\ge3/L$ for $L\ge6$,
Eq.~\eqref{eq:ringenergy} yields the sharp bound
\begin{equation}
E_0(L,\varphi)
\ge
-\frac{2t_{\rm DW}}{3}L,
\label{eq:loopbound}
\end{equation}
with equality if and only if $(L,\varphi)=(6,-1)$; the lower-bound derivation is given in Appendix~\ref{app:ring}.  Thus an
antiperiodic elementary hexagon uniquely minimizes the ground-state energy per
loop site.

The excess energy associated with a nonoptimal loop length or twist is defined by
\begin{equation}
\Delta E_\ell
\equiv
E_0(L_\ell,\varphi_\ell)
+
\frac{2t_{\rm DW}}{3}L_\ell.
\label{eq:deltaE}
\end{equation}
Eq.~\eqref{eq:loopbound} implies
$\Delta E_\ell\ge0$, and $\Delta E_\ell=0$ for $(L_\ell,\varphi_\ell)=(6,-1)$.
Summing over all loops, the unprojected matter energy becomes
\begin{equation}
E_{\rm M}^{(0)}
=
-\frac{2t_{\rm DW}}{3}N_{\rm act}
+
\sum_\ell\Delta E_\ell.
\label{eq:E_Nact}
\end{equation}
Here $N_{\rm act}$ counts both the active Majorana bonds and the vertices of the loop network, since each active vertex has degree two. 
Eq.~\eqref{eq:E_Nact} separates two energetic contributions: the first favors a
maximal number of active bonds, whereas the second penalizes loops whose
length or Wilson-loop twist differs from the optimal
$(L,\varphi)=(6,-1)$ configuration.

The first contribution can be written entirely in terms of the flux
variables.  A bond of the dual triangular lattice is active precisely
when $w_p\neq w_q$, so
\begin{equation}
N_{\rm act}
=
\frac{1}{2}
\sum_{\langle pq\rangle}
\left(1-w_pw_q\right).
\label{eq:Nact_Ising}
\end{equation}
Since the $N_p$-site triangular lattice contains $3N_p$ nearest-neighbor
bonds,
\begin{equation}
\sum_{\langle pq\rangle}w_pw_q
=
3N_p-2N_{\rm act}.
\label{eq:Ising_identity}
\end{equation}
Substitution into Eq.~\eqref{eq:E_Nact} yields the exact identity
\begin{equation}
E_{\rm M}^{(0)}
=
-t_{\rm DW}N_p
+
\frac{t_{\rm DW}}{3}
\sum_{\langle pq\rangle}w_pw_q
+
\sum_\ell\Delta E_\ell.
\label{eq:ising_decomposition}
\end{equation}
Up to a constant, the second term above is precisely
the energy of a nearest-neighbor antiferromagnetic Ising model on the triangular
lattice with coupling $J_{\rm eff}=t_{\rm DW}/3>0$.
Its ground-state constraint requires exactly one frustrated
equal-sign bond on every elementary triangle.  As discussed in
Sec.~\ref{sec:loops}, this is equivalent to a
fully packed Majorana-loop configuration.  The Ising part thus
selects an extensively degenerate fully packed manifold from the full
set of flux sectors, subject to the global constraint
$\prod_p w_p=1$ on the torus.

The residual term $\sum_\ell\Delta E_\ell$ then distinguishes states
within this manifold.  Since $\Delta E_\ell$ is nonnegative,
the Ising lower bound can be saturated only if every loop is an
elementary hexagon with $\varphi_\ell=-1$.  
The two terms thus play complementary roles: the Ising
contribution selects the degenerate fully packed manifold, and the Majorana
zero-point energy selects among its flux configurations according to both loop
geometry and twist.

Two points clarify the status of this decomposition. First,
Eq.~\eqref{eq:ising_decomposition} is an identity, not an approximation: the
reference energy $-2t_{\rm DW}/3$ per active site in Eq.~\eqref{eq:E_Nact} is
fixed uniquely by requiring the residual to be nonnegative and tight, and any
other choice merely redistributes the same total energy between the two terms.
Second, flux selection follows from the inequality of
Sec.~\ref{sec:groundstate}, not from the Ising rewriting, which only exposes
the competition between maximizing the active-bond count and optimizing loop
geometry and twist. The mechanism is a nonperturbative counterpart of
fermion-mediated order by
disorder~\cite{Villain1980,Shender1982,Chalker1992}: fermionic zero-point
energy lifts an extensive classical degeneracy with no small parameter,
scale separation, or fluctuation expansion.

\section{Exact flux selection at the loop point}
\label{sec:groundstate}

The flux configuration minimizing the unprojected Majorana ground-state energy at the loop point is determined as follows. 
Up to and including Sec.~\ref{subsec:uniqueness}, $E_{\rm M}^{(0)}$
denotes the unprojected energy of the free-Majorana problem.  Whether the
minimizer survives the gauge projection onto the physical spin Hilbert space is
a separate question whose answer depends on the sign of
$J_0$; we address it in Sec.~\ref{subsec:physical} and
Appendix~\ref{app:finite-torus-ed} before drawing any conclusion about the spin
model.  The exact decomposition
of Eq.~\eqref{eq:E_Nact} makes the minimization particularly transparent.  Since
$N_{\rm act}\le N_s$ and $\Delta E_\ell\ge0$ for every loop,
\begin{equation}
E_{\rm M}^{(0)}
\ge
-\frac{2t_{\rm DW}}{3}N_{\rm act}
\ge
-\frac{2t_{\rm DW}}{3}N_s .
\label{eq:globalbound}
\end{equation}
The second inequality is saturated only when every honeycomb site
belongs to the active loop network, $N_{\rm act}=N_s$, and the
first requires $\Delta E_\ell=0$ for every loop.
The latter condition is equivalent to
$(L_\ell, \varphi_\ell)=(6, -1)$ for every loop.
Thus a configuration saturates the global lower bound if and only if
the honeycomb lattice is fully covered by disjoint antiperiodic
elementary Majorana hexagons.
 
\begin{figure}[t]
\centering
\includegraphics[width = \linewidth]{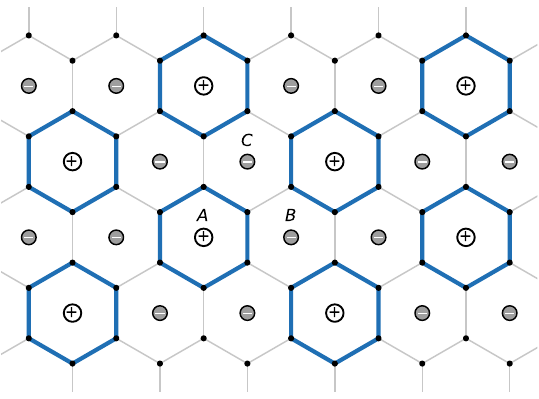}
\caption{Flux configuration of the $2/3$-vison crystal.  
Flux-free $A$ plaquettes (circled $+$) are surrounded
by visons on the $B$ and $C$ sublattices (filled circles, $w_p=-1$), and their
boundaries form disjoint elementary Majorana rings covering every site.  On
admissible commensurate tori, this pattern is unique up to the three
translations of the flux-free sublattice.
}\label{fig:crystal}
\end{figure}

A flux configuration satisfying these conditions is obtained from the
three-sublattice decomposition of the dual triangular lattice.  Let
$A$, $B$ and $C$ denote sublattices and 
\begin{equation}
w_A=+1, \qquad w_B=w_C=-1.
\label{eq:23flux}
\end{equation}
Each $A$ plaquette is surrounded by three $B$ and three $C$ plaquettes and thus has six active boundary bonds [see Fig.~\ref{fig:crystal}].  Every honeycomb vertex
touches exactly one plaquette of each sublattice.  Consequently, the
boundaries of the $A$ plaquettes are mutually vertex-disjoint and
cover every honeycomb site exactly once.  The active Majorana graph is
therefore a close-packed array of elementary hexagonal loops, so that
$N_{\rm act}=N_s$.
 
Moreover, each such loop encloses an $A$ plaquette with $w_A=+1$, 
and $\varphi_{\rm hex}=-w_A=-1$.
Thus every hexagon also saturates the single-loop bound, and
this flux pattern saturates the global bound exactly:
\begin{equation}
\frac{E_{\rm M}^{(0)}}{N_s}
=
-\frac{2t_{\rm DW}}{3}
=
-\frac{4|J_0|}{3}.
\label{eq:estar}
\end{equation}
This establishes the state as an exact minimizer of the unprojected
matter energy, rather than as the outcome of a variational comparison.
On admissible commensurate tori,
the number of visons is $2N_p/3$, which is even, and the global constraint $\prod_p w_p=1$
is therefore satisfied.

\subsection{Uniqueness of the saturating local-flux pattern}
\label{subsec:uniqueness}

We now determine whether any other local flux configurations can
saturate Eq.~\eqref{eq:globalbound}.  Saturation requires $N_{\rm act}=N_s$ and
$(L_\ell,\varphi_\ell)=(6,-1)$ for every loop, so each honeycomb vertex
belongs to exactly one six-site loop.  Every contractible six-cycle of the
honeycomb lattice bounds a single plaquette, which must be flux-free by
Eq.~\eqref{eq:loop_twist_flux}.  The only alternative, a noncontractible
six-cycle, is excluded by the systole condition of Sec.~\ref{sec:loops}.
The classification below is therefore exhaustive.

To implement this constraint on the dual triangular lattice, 
we label its sites by integer coordinates $(m,n)$ along two primitive
directions, with
\begin{equation}
\eta_{m,n}
=
\frac{1+w_{m,n}}{2}
=
\begin{cases}
1, & w_{m,n}=+1,\\
0, & w_{m,n}=-1.
\end{cases}
\end{equation}
Because every honeycomb vertex belongs to exactly one saturating hexagon, the
corresponding elementary dual triangle contains exactly one site with $\eta=1$.
For the two triangle orientations,
\begin{equation}
\begin{aligned}
\eta_{m,n}+\eta_{m+1,n}+\eta_{m,n+1}
&=1,\\
\eta_{m+1,n}+\eta_{m,n+1}+\eta_{m+1,n+1}
&=1.
\end{aligned}
\label{eq:triangle_constraints}
\end{equation}
Subtracting the two constraints gives
$\eta_{m,n}=\eta_{m+1,n+1}$, so $\eta_{m,n}$ depends only on
$d=m-n$: $\eta_{m,n}=g_d$.  Then the triangle constraint becomes
\begin{equation}
g_{d-1}+g_d+g_{d+1}=1.
\end{equation}
Subtracting the equation at $d+1$ from that at $d$ gives
$g_{d-1}=g_{d+2}$, or $g_{d+3}=g_d$.
Since $g_d\in\{0,1\}$ and every three consecutive entries sum to one, the
only solutions are the three translations of
$\ldots100100100\ldots$.  Therefore, on tori commensurate with the
three-sublattice order and satisfying the systole
condition, only the three translations of Eq.~\eqref{eq:23flux} saturate the
exact matter-energy bound.  

The resulting state is therefore a translation-broken \emph{hexagonal
Majorana-loop crystal}, whose matter Hamiltonian decomposes into gapped
antiperiodic six-site rings.
On a torus, local plaquette fluxes leave the two noncontractible Wilson loops unfixed. Since every active component is a contractible hexagon, the ring twists and unprojected Majorana spectrum are independent of global holonomies, yielding four degenerate Wilson sectors at the quadratic level. This fourfold degeneracy is distinct from the three translation branches and, although consistent with intrinsic $\Ztwo$ topological order, does not by itself establish it: at the loop point, holonomy independence follows automatically from the contractible-ring decomposition.

\subsection{Physical ground state after gauge projection}
\label{subsec:physical}

Projection onto the physical spin Hilbert space imposes an additional matter-parity constraint through $P_{\rm phys}=\prod_i(1+D_i)/2$. 
Consequently, the unprojected bound is attained by the spin model only when the corresponding quasiparticle vacuum survives this projection.   
Appendix~\ref{app:finite-torus-ed} derives the surviving matter parity and
evaluates it in the crystal sector.  
With $s=\sgn(J_0)$ and $N_{\rm hex}=N_s/6$ denoting the number of rings,
\begin{equation}
\begin{aligned}
P_f^{\rm phys}&=(-s)^{N_{\rm hex}},\\
E_{\rm spin}^{\rm cr}-E_{\rm M}^{(0)}&=t_{\rm DW}\bigl[1-(-s)^{N_{\rm hex}}\bigr],
\end{aligned}
\label{eq:parityrule_main}
\end{equation}
independently of the Wilson sector.  
Here $P_f^{\rm phys}=(-1)^{N_f}$ is the parity of the matter-fermion number that
survives projection---not a Wilson-loop eigenvalue or a vison count---and
$E^{\rm cr}_{\rm spin}$ is the lowest physical energy within the crystal flux
sector, which need not be the ground-state energy of the full spin Hamiltonian.
This has been verified against direct spin
exact diagonalization on six commensurate tori, $N_s=18,\ldots,48$.

The parity rule has two distinct consequences.  For antiferromagnetic coupling $J_0<0$,
$P_f^{\rm phys}=+1$ for any $N_{\rm hex}$; for ferromagnetic coupling $J_0>0$, the same holds when
$N_{\rm hex}$ is even.  In either case, the crystal vacuum is physical and its
projected energy attains the absolute all-sector bound,
$E_{\rm spin}^{\rm cr}=-\frac{2}{3}t_{\rm DW}N_s$,
proving that the $2/3$-vison crystal is a ground state of the spin
Hamiltonian~\eqref{eq:HDWspin}.  Because the matter Hamiltonian is a direct
sum of vertex-disjoint contractible rings, its spectrum is exactly holonomy
independent.  The three translations and four holonomy sectors therefore
give an exact twelvefold degeneracy at every such finite commensurate size,
without finite-size splitting.  Appendix~\ref{app:local-indistinguishability}
independently recovers the fourfold factor by mapping the crystal sector,
through a depth-one local unitary, onto a toric-code stabilizer model with two
logical qubits.

For ferromagnetic coupling $J_0>0$ and odd $N_{\rm hex}$, the vacuum is
rejected and the lowest physical state of the crystal sector carries one matter
quantum, raising its energy by $2t_{\rm DW}$.  This correction is $O(1)$ and does
not affect the thermodynamic energy density, but it can reorder sectors at finite
size: on the $N_p=9$ torus, an exhaustive search over all flux and Wilson sectors
places the physical minimum in a single $L=18$ loop sector rather than in the $2/3$-vison
crystal (for details see Appendix~\ref{app:finite-torus-ed}).  
Therefore, Eq.~\eqref{eq:globalbound} holds for the unprojected matter energy on any
admissible commensurate torus, while its saturation by the physical crystal requires
$P_f^{\rm phys}=+1$.

\subsection{Domain walls and incommensurate tori}
\label{subsec:walls}

Two questions remain after the uniqueness argument: the energy cost of a
boundary between crystal translations and the behavior on tori
incommensurate with the three-sublattice order.  Using
$N_{\rm act}=N_s-N_0$, Eq.~\eqref{eq:E_Nact} gives the exact excess above the
absolute bound,
\begin{equation}
\delta E
\equiv
E_{\rm M}^{(0)}+\frac{2t_{\rm DW}}{3}N_s
=
\frac{2t_{\rm DW}}{3}N_0+\sum_\ell\Delta E_\ell .
\label{eq:excess_energy}
\end{equation}
Contractible loops have lengths $L=6,10,12,14,\ldots$.  Their smallest
nonzero excess is
\begin{equation}
\begin{aligned}
\Delta E_{\min}^{\rm con}
&\equiv
\min_{\substack{\text{contractible}\\(L,\varphi)\neq(6,-1)}}
\Delta E(L,\varphi) \\
&=\Delta E(10,-1)
\simeq0.195\,t_{\rm DW}.
\end{aligned}
\label{eq:deltaEmin}
\end{equation}
An admissible torus may also support noncontractible eight-cycles, for which
$\Delta E(8,-1)\simeq0.107\,t_{\rm DW}$.  Since the antiperiodic branch is
lowest and $\Delta E(L,-1)/L$ increases with $L$, every nonoptimal loop on an
admissible torus obeys
\begin{equation}
\frac{\Delta E(L,\varphi)}{L}
\ge
\epsilon_{\rm loop}
\equiv
\frac{\Delta E(8,-1)}{8}
\simeq0.0134\,t_{\rm DW}.
\label{eq:loop_excess_density}
\end{equation}
Define the defect support by
\begin{equation}
S_{\rm def}
\equiv
N_0+
\sum_{\ell:\,(L_\ell,\varphi_\ell)\neq(6,-1)}L_\ell,
\label{eq:defect_support}
\end{equation}
which counts the honeycomb sites not belonging to optimal antiperiodic
hexagonal loops.  Equations~\eqref{eq:excess_energy}
and~\eqref{eq:loop_excess_density} imply
$\delta E\ge\epsilon_{\rm loop}S_{\rm def}$.
A boundary between distinct crystal translations cannot be composed entirely
of optimal hexagons and must therefore have $S_{\rm def}>0$.  This establishes
a positive local defect scale, but not by itself a positive domain-wall
tension.  The latter additionally requires a size-independent constant $c>0$
such that $S_{\rm def}\ge c\ell_w$ for a wall of length $\ell_w$.  If this
geometric counting bound holds, then
\begin{equation}
\delta E\ge c\epsilon_{\rm loop}\ell_w\equiv\tau\ell_w,
\qquad \tau>0.
\label{eq:conditional_wall_tension}
\end{equation}
At low temperature the Boltzmann weight of an individual wall is then
suppressed approximately as $\exp(-\tau\ell_w/T)$.
Establishing the required counting
bound, and determining whether wall proliferation produces a finite-temperature
transition, are left for future work.

On a fixed-aspect-ratio torus incommensurate with the three-sublattice order,
exact saturation is impossible.  One can choose a trial flux configuration
that agrees with a crystal translation except within a noncontractible strip
of fixed width.  Only $O(L)$ sites then belong to defective components, and
since $0\le\Delta E_\ell\le(2t_{\rm DW}/3)L_\ell$, the trial-state excess is
at most $O(L)$.  Denoting by
$E_{\min}^{(0)}(L)$ the minimum unprojected matter energy on this torus, the
global lower bound and this variational construction give
\begin{equation}
0\le
\frac{E_{\min}^{(0)}(L)}{N_s}
+\frac{2t_{\rm DW}}{3}
\le\frac{C}{L},
\label{eq:incommensurate_bound}
\end{equation}
where $C$ is independent of $L$.  Thus commensurability is required for exact
finite-size saturation, but not for the thermodynamic energy density of
Eq.~\eqref{eq:estar}.

\subsection{Flux defects and loop reconnection}
\label{subsec:defects}

The loop representation makes local flux rearrangements transparent.
Flipping a flux-free $A$ plaquette of the selected $2/3$-vison crystal,
$w_A:+1\to-1$, removes its antiperiodic hexagonal ring and isolates its six
boundary vertices.  The resulting six Majorana zero modes form three complex
zero modes, giving an eight-dimensional zero-mode Fock space before
projection, while the unprojected matter energy increases by
$\delta E_A=4t_{\rm DW}$.  When this defect is embedded in an allowed flux
configuration, the global parity constraint selects the physical subspace.

By contrast, flipping a $B$ or $C$ plaquette, $w_{B/C}:-1\to+1$, replaces
three active bonds by three others.  Thus $N_{\rm act}$ is unchanged and three
neighboring hexagons reconnect into one contractible antiperiodic $L=18$ loop.
Its unprojected energy cost is
\begin{align}
\delta E_{B/C}=
E_0(18,-)-3E_0(6,-)
\simeq
0.4825\,t_{\rm DW}.
\label{eq:BCdefect}
\end{align}
This reconnection produces no zero modes and costs roughly eight times less
than the $A$-plaquette flip.

The loop decomposition also determines the exact zero-mode count.  Each
isolated site contributes one Majorana zero mode.  Since the bipartite
honeycomb lattice supports only even-length loops, each periodic loop
($\varphi_\ell=+1$) contributes two additional zero modes, at $k=0$ and
$k=\pi$, whereas an antiperiodic loop contributes none.  Hence
\begin{equation}
N_{\rm zero}=N_0+2N_{\rm P},
\label{eq:zero-mode-count}
\end{equation}
where $N_{\rm P}$ is the number of periodic loops.  Thus zero modes are
controlled jointly by connectivity and twist, rather than by flux defects alone.

The crystal-sector single-particle gap is
\begin{equation}
\Delta_{\rm sp}
=\min_{\epsilon_n>0}\epsilon_n
=2t_{\rm DW},
\label{eq:gap_sp}
\end{equation}
but it is not a physical spin gap.  For $J_0<0$, the Majorana vacuum is
physical and the parity constraint requires fixed-sector excitations to
occupy at least two modes, giving
\begin{equation}
\Delta_{\rm sec}=4t_{\rm DW}.
\label{eq:gap_phys}
\end{equation}
By contrast, the local unprojected flux-defect costs are
$\delta E_A=4t_{\rm DW}$ and
$\delta E_{B/C}\simeq0.4825\,t_{\rm DW}$.  Neither is the physical flux gap
on a closed torus, where $\prod_p w_p=1$ restricts flux changes to allowed
multiplaquette combinations with generally nonadditive energies.

Nevertheless, the full spin gap has a size-independent lower bound.
On an admissible commensurate torus, the result of
Sec.~\ref{subsec:uniqueness} implies that every sector outside the crystal
ground-state manifold contains an isolated site or a nonoptimal loop.
Equation~\eqref{eq:excess_energy} then places its unprojected vacuum at least
$\min\{2t_{\rm DW}/3,\Delta E(8,-1)\}=\Delta E(8,-1)$ above the ground energy.
Projection cannot reduce this energy, whereas excitations within a crystal
ground-state block cost at least $\Delta_{\rm sec}=4t_{\rm DW}$.  Therefore
\begin{equation}
\Delta_{\rm spin}\ge\min\{\Delta E(8,-1),4t_{\rm DW}\}
\simeq0.107t_{\rm DW}.
\label{eq:full_spin_gap_bound}
\end{equation}
If noncontractible eight-cycles are absent, the bound improves to
$\Delta E(10,-1)\simeq0.195\,t_{\rm DW}$.  These are lower bounds, not exact
spin gaps, but they establish a size-independent spectral gap under the
stated geometric conditions.

\section{Toric-code structure of the vison crystal}
\label{sec:toriccode}

\begin{figure*}[t]
\centering
\includegraphics[width=\textwidth]{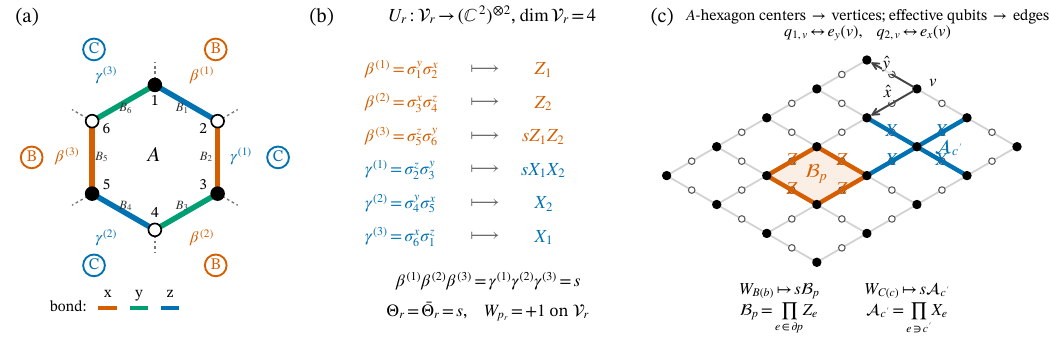}
\caption{Depth-one local-unitary construction.
(a) A flux-free $A$ hexagon of the $2/3$-vison crystal. 
The dashed stubs mark inactive $--$ bonds. The six neighboring plaquettes alternate between $B$ and $C$. Their respective two-spin factors within the hexagon are $\beta_r^{(a)}$ and $\gamma_r^{(a)}$ [Eq.~\eqref{eq:app-betagamma}].
(b) These six factors lie in the commutant of the hexagon bond algebra and restrict to a two-qubit Pauli frame on the four-dimensional cluster ground space $\mathcal V_r$ [Eq.~\eqref{eq:app-pauliframe}]. The implementing unitary $U_r$ acts on the six spins of one hexagon; consequently, $U=\bigotimes_rU_r$ has depth one when each six-spin unitary is counted as one local gate.
(c) Edge relabeling. Each $B$-plaquette stabilizer maps to the weight-four $Z$ boundary $\mathcal B_p$ of an effective square face, and each $C$-plaquette stabilizer to the weight-four $X$ star $\mathcal A_{c'}$ at an effective vertex. The effective square lattice is generally sheared in the physical embedding; the mapping is combinatorial.
}\label{fig:toricmap}
\end{figure*}

The physical crystal ground space is characterized first
on the antiferromagnetic branch $J_0<0$, for which the
matter vacuum survives projection at every admissible
commensurate size.  The same construction applies to $J_0>0$ when
$N_{\rm hex}$ is even; the incompatible odd case is treated separated below.
The factorization into vertex-disjoint hexagons makes the structure
analytic: a depth-one local unitary maps the crystal ground space directly onto
the ground space of a square-lattice toric code.  
The construction and its consequences are summarized here.
The full hexagon algebra, signs, and coordinate
bookkeeping are supplied in Appendix~\ref{app:local-indistinguishability}, with
numerical checks reported in Appendix~\ref{app:ed-check} and \ref{app:strings}.

Kitaev bond operators commute with every plaquette operator, $[B_e,W_p]=0$, so
within the crystal flux sector each $W_p$ in Eq.~\eqref{eq:HDWspin} may be
replaced by its eigenvalue.  
The factor $1-W_{p_e}W_{q_e}$ consequently equals $2$ on active bonds and $0$ elsewhere, which gives
\begin{equation}
 H_{\rm DW}^{\rm spin}\Big|_{\rm crystal}
 =-2J_0\sum_{r=1}^{N_{\rm hex}}\;\sum_{e\in\partial p_r}B_e
 \;\equiv\;\sum_{r=1}^{N_{\rm hex}} h_r ,
 \label{eq:cluster-form}
\end{equation}
where $p_r$ runs over the flux-free $A$ plaquettes.  These hexagons are mutually
vertex-disjoint and every active bond belongs to exactly one of them.
Hence the $h_r$ act on disjoint sets of six spins and commute with each other.
However, for ground-space construction, we first regard
the right-hand side of Eq.~\eqref{eq:cluster-form} as an auxiliary Hamiltonian on the unrestricted spin
Hilbert space and impose the crystal flux constraints below.

Let $\mathcal H_r=(\mathbb C^2)^{\otimes6}$ denote the six-spin Hilbert
space of cluster $r$, and let
$\mathcal A_r=\operatorname{alg}\{B_1,\ldots,B_6\}$ be the algebra generated
by its six bond operators~\cite{Nussinov2009,Cobanera2011}. 
As shown in Fig.~\ref{fig:toricmap}(a), $B_j$ joins sites $j$ and $j+1$ modulo $6$.
The alternating products
$\Theta_r=B_1B_3B_5$ and $\bar\Theta_r=B_2B_4B_6$ commute with every generator
and satisfy $\Theta_r\bar\Theta_r=W_{p_r}$, so the center of $\mathcal A_r$ is
spanned by $\{1,\Theta_r,\bar\Theta_r,W_{p_r}\}$. 
These central elements are products of bond operators on disjoint site pairs and are therefore traceless,
so their four characters split $\mathcal H_r$ into sectors of equal dimension
$64/4=16$.  Removing the two central combinations from the six generators
leaves four independent ones, which generate a four-dimensional irreducible
representation; hence $\mathcal A_r\cong M_4(\mathbb C)^{\oplus4}$, and since
$16=4\times4$ that representation occurs with multiplicity four in every sector.
Evaluating $h_r$ in the four blocks places its minimum in the sector
$\Theta_r=\bar\Theta_r=\sgn(J_0)$, nondegenerate within the corresponding
$M_4(\mathbb C)$ factor; the remaining fourfold multiplicity is the cluster ground space,
\begin{equation}
 \dim\mathcal V_r=4,
 \qquad
 h_r\big|_{\mathcal V_r}=-4t_{\rm DW}\,\mathbb I_{\mathcal V_r}.
 \label{eq:Vr-main}
\end{equation}
In particular $W_{p_r}=\Theta_r\bar\Theta_r=+1$ on $\mathcal V_r$, so the
flux-free $A$-plaquette condition is met by every cluster ground state rather
than imposed.
 
Writing $P_r$ for the projector onto $\mathcal V_r$, every
$a\in\mathcal A_r$ satisfies $P_raP_r=c_aP_r$ with $c_a\in\mathbb C$: the bond
algebra cannot resolve the fourfold multiplicity, which instead encodes two
effective qubits.  Their operator algebra is supplied by the commutant
$\mathcal A_r'=\{O:[O,a]=0\ \forall a\in\mathcal A_r\}$ rather than by $\mathcal A_r$
itself~\cite{Knill2000,Zanardi2001}, and compressing it onto the ground
space returns the full two-qubit algebra, $P_r\mathcal A_r'P_r\cong M_4(\mathbb C)$.
Recovering the prescribed crystal sector within
$\bigotimes_r\mathcal V_r$ still requires imposing the neighboring $B$- and
$C$-plaquette constraints.  Their two-spin factors inside hexagon $r$, denoted
$\beta_r^{(a)}$ and $\gamma_r^{(a)}$ in Fig.~\ref{fig:toricmap}(a), commute
with all six bond operators and hence belong to $\mathcal A_r'$; they preserve
$\mathcal V_r$, where their projected actions furnish the $Z$- and $X$-type
Paulis of the two effective qubits.
 
Let $U_r$ be a unitary on the six spins of hexagon $r$ that maps $\mathcal V_r$
onto $(\mathbb C^2)^{\otimes2}\otimes|0\rangle^{\otimes4}$ and implements that
Pauli frame, and set $U=\bigotimes_{r=1}^{N_{\rm hex}}U_r$ [see Fig.~\ref{fig:toricmap}(b)].
Because the hexagons are vertex disjoint, $U$ is a local unitary circuit of
depth one whose support radius is a single hexagon.
Under $U$ the $A$-plaquette constraints are automatically satisfied.  Each
remaining $B$ or $C$ plaquette meets three neighboring $A$ hexagons, but its
effective Pauli weight is \emph{four}, not three.

As shown in Fig.~\ref{fig:toricmap}(c), the two effective qubits $q_{2,v}$ and
$q_{1,v}$ on each $A$ hexagon are assigned respectively to the $\hat x$- and
$\hat y$-directed edges of a sheared square cellulation whose vertices are the
$A$-hexagon centers, with $\hat x$ and $\hat y$ primitive vectors of the
triangular lattice they form.  With the plaquettes relabeled accordingly,
\begin{equation}
\begin{aligned}
 U W_{B(b)}U^\dagger\big|_{\rm eff}&=s\,\mathcal B_p,
 &\quad \mathcal B_p&=\prod_{e\in\partial p}Z_e,\\
 U W_{C(c)}U^\dagger\big|_{\rm eff}&=s\,\mathcal A_{c'},
 &\quad \mathcal A_{c'}&=\prod_{e\ni c'}X_e ,
\end{aligned}
\label{eq:toric-checks-main}
\end{equation}
the weight-four plaquette and star operators.  The explicit coordinate
relabeling is given in Appendix~\ref{app:toric-explicit}.  The effective
constraint algebra is therefore exactly that of the toric code on this
cellulation, not a code inferred to lie in the same
phase~\cite{BombinDuclosPoulin2012,Bombin2014}.

The fourfold code-space degeneracy follows directly from the torus topology.  The
effective cellulation has $V=F=N_{\rm hex}$ and $E=2N_{\rm hex}$ edge
qubits.  Each stabilizer family has one global redundancy,
\begin{equation}
\prod_{p\in B}W_p=\prod_r\bar\Theta_r,
\qquad
\prod_{p\in C}W_p=\prod_r\Theta_r,
\label{eq:two-relations}
\end{equation}
where the right-hand sides are fixed scalars on
$\bigotimes_r\mathcal V_r$.  Thus only $(V-1)+(F-1)$ stabilizers are
independent, leaving
\begin{equation}
k=E-(V-1)-(F-1)=2,
\quad
\dim\mathcal G=2^k=4.
\label{eq:code-dim}
\end{equation}
The two logical qubits are associated with Wilson loops around the two
noncontractible cycles of the torus.

Compatibility with the physical projection imposes an
additional sign constraint on this four-dimensional code space.
The crystal has $w_B=w_C=-1$, so Eq.~\eqref{eq:toric-checks-main} and
$s^2=1$ fix the required eigenvalues,
\begin{equation}
 \mathcal B_p=\mathcal A_{c'}=-s
 \qquad\text{for every }p\text{ and }c' .
 \label{eq:toric-required-stabilizers}
\end{equation}
Each family has $N_{\rm hex}$ members, by the counting above.
On a torus the plaquette and star stabilizers $\mathcal B_p$
and $\mathcal A_{c'}$ are not independent but obey
the standard redundancies $\prod_p\mathcal B_p=\prod_{c'}\mathcal
A_{c'}=\mathbb I$, so multiplying Eq.~\eqref{eq:toric-required-stabilizers} over all
plaquettes gives the compatibility condition
\begin{equation}
 (-s)^{N_{\rm hex}}=1 .
 \label{eq:toric-sign-compatibility}
\end{equation}
Equivalently, the crystal is compatible for either sign of $J_0$ when
$N_{\rm hex}$ is even, and for $s=-1$ alone when $N_{\rm hex}$ is odd.
This is not the honeycomb flux constraint in disguise: $\prod_pW_p=1$ holds
identically in the crystal sector, the $A$ sublattice contributing
$(+1)^{N_{\rm hex}}$ and the $B$ and $C$ sublattices $(-1)^{N_{\rm hex}}$ each.

Equation~\eqref{eq:toric-sign-compatibility} establishes the projection rule of
Eq.~\eqref{eq:parityrule_main} at general commensurate size.  The two are one
statement in two languages: a property of the stabilizer group on a closed
surface, involving no fermions, and its expression through the gauge constraint
$\prod_iD_i=1$ of the Majorana representation.  Independent confirmation comes
from the spin exact diagonalization of Appendix~\ref{app:finite-torus-ed},
which matches the projected spectra on six commensurate tori.
The three cases are then transparent.  For $J_0<0$ all stabilizers equal $+1$:
the crystal maps onto the unexcited toric-code ground space at every
commensurate size, and its four holonomy states, together with the three
translated crystals, give the twelvefold count of
Sec.~\ref{subsec:physical}.  For $J_0>0$ all stabilizers equal $-1$, so every
plaquette and every star carries an excitation; with $N_{\rm hex}$ even this
all-$(-1)$ sector is related to the conventional one by a product of
single-edge Paulis, whereas with $N_{\rm hex}$ odd
Eq.~\eqref{eq:toric-sign-compatibility} fails, since a closed surface admits
each species only in pairs.  No state in $\bigotimes_r\mathcal V_r$ then lies in
the physical crystal sector, and the lowest physical level must leave one
cluster ground space at cost $\epsilon_{\min}=2t_{\rm DW}$, to which the
four-state mapping does not apply.

Nontrivial logical Pauli operators are noncontractible cycles of the primal or
dual effective square lattice, so the effective code distance is $d_{\rm eff}=\min(\ell_{\rm sys},\ell_{\rm sys}^{*})$,
with $\ell_{\rm sys}$ and $\ell_{\rm sys}^{*}$ the corresponding systoles.  The
depth-one unitary transfers this local indistinguishability to the spin ground
space up to a one-hexagon thickening of support, and the Wilson loops and
conjugate dual strings of Appendix~\ref{app:strings} realize the two logical
Pauli pairs.  Because the matter sector is a time-reversal-invariant atomic
limit of decoupled rings, it has vanishing Chern number and lies in the $\nu=0$
class of Kitaev's sixteenfold way~\cite{Kitaev2006,ZhangPRR2020}.
On the compatible branch the selected state is thus a toric-code ground state
dressed by a six-spin depth-one unitary, coexisting with broken translation
symmetry.

The exact ground-space mapping is restricted to the fine-tuned loop point.
Away from it the plaquette fluxes remain conserved, but the matter Hamiltonian
no longer decomposes into disconnected contractible loops, so neither the loop
bound nor the proof of vison-crystal selection applies.  Fluxes become dynamical
only under perturbations that fail to commute with $W_p$; whether the crystal
and its topological properties survive these is left for future work.

\section{Away from the loop point}
\label{sec:away}

\subsection{Fixed-sector matter gap}

We first examine the matter spectrum while keeping the $2/3$-vison flux
pattern of Eq.~\eqref{eq:23flux} fixed.  This sector contains no $++$
bonds; defining $t_d\equiv |J_{+-}|$ and $\delta\equiv J_{--}$,
the loop point has $\delta=0$ and $t_d=t_{\rm DW}$.
More generally, setting $\delta=0$ already suffices to decouple the Majorana
hexagons within this sector.  In the notation of Eq.~\eqref{eq:gap_sp}, the
corresponding single-particle gap is $\Delta_{\rm sp}=2t_d$,
which reduces to Eq.~\eqref{eq:gap_sp} at the loop point.  

For finite $\delta$, the $--$ bonds couple neighboring hexagons and form a
perfect matching.  In the Hermitian single-particle matrix $\ii A$, each such
bond contributes a $2\times2$ block with eigenvalues $\pm2|\delta|$.  Since
the matching bonds do not share vertices, the full perturbation has operator
norm $2|\delta|$.  Weyl's inequality therefore gives
\begin{equation}
\Delta_{\rm sp}
\ge
2\bigl(t_d-|\delta|\bigr).
\label{eq:gapbound}
\end{equation}
Hence the fixed-sector single-particle gap remains open whenever $|J_{--}|<|J_{+-}|$.

This bound is a statement about the spectrum of a \emph{fixed} flux sector,
not about the stability of the $2/3$-vison crystal.  In particular, it contains
no information about the energies of competing flux configurations and is
independent of $J_{++}$, which does not appear in the $2/3$-vison sector.  At
the pure Kitaev point $J_1=J_2=0$, for example, the ground state belongs to
the zero-flux sector, while Eq.~\eqref{eq:gapbound} is only marginal.
Flux-sector competition must therefore be addressed separately.

\begin{table}[t]
\centering
\setlength{\tabcolsep}{4pt}
\begin{tabular}{lcl}
\hline\hline
State & \quad $\rho_v$ & \quad Local flux pattern \\
\hline
$2/3$ vison & \quad $2/3$ & \quad $(+,-,-)$ three-sublattice \\
$1/3$ vison & \quad $1/3$ & \quad $(-,+,+)$ three-sublattice \\
$1/4$ vison & \quad $1/4$ & \quad $w_{2m,2n}=-1$, otherwise $+1$  \\
$3/4$ vison & \quad $3/4$ & \quad flux complement of $1/4$-vison \\
Zero flux & \quad $0$ & \quad $w_p=+1$ for all $p$ \\
$\pi$ flux & \quad $1$ & \quad $w_p=-1$ for all $p$ \\
$1/2$-vison stripe & \quad $1/2$ & \quad $w_{m,n}=(-1)^m$, otherwise $+1$ \\
\hline\hline
\end{tabular}
\caption{Seven representative periodic flux patterns used in the finite-size
candidate-sector comparison (see Fig.~\ref{fig:sectorcuts}).  Here $\rho_v$ denotes the vison density.}
\label{tab:flux_patterns}
\end{table}

\begin{figure}[t]
\centering
\includegraphics[width=\linewidth]{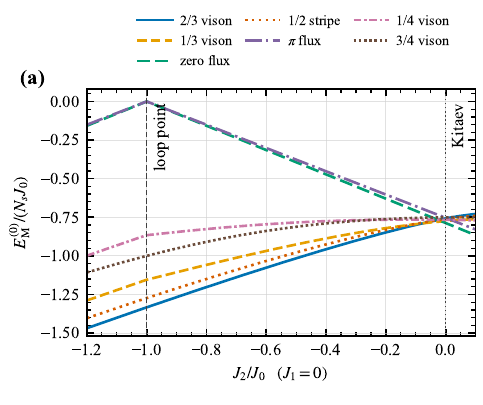}
\includegraphics[width=\linewidth]{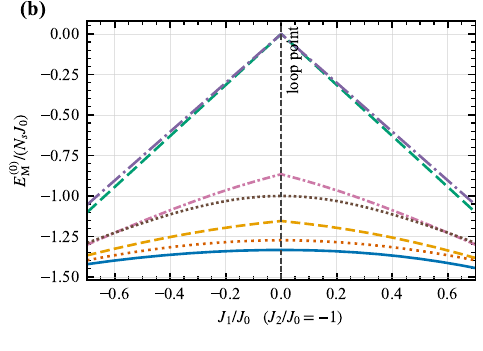}
\caption{Unprojected matter ground-state energies per spin, in units of $J_0$, for seven representative periodic flux sectors on a $120\times120$-unit-cell torus, 
minimized over the four global Wilson-loop sectors.
(a) Along $J_1=0$, the $2/3$-vison sector is favored near the loop point, while the flux-free sector becomes lowest toward the pure Kitaev limit.
(b) Along $J_2/J_0=-1$, the $2/3$-vison sector remains lowest throughout; the loop point lies at $J_1/J_0=0$.
}\label{fig:sectorcuts}
\end{figure}

\subsection{Representative flux-sector competition}
\label{sec:candidate}

To probe flux-sector competition, we compare seven
symmetry-inequivalent periodic flux patterns corresponding to those
considered in Ref.~\cite{ZhangPRL2019} (see Table~\ref{tab:flux_patterns}): the
uniform zero- and $\pi$-flux states, the three-sublattice $1/3$- and
$2/3$-vison states, a representative $1/2$-vison stripe (one of three
$C_3$-related orientations), and complementary four-sublattice $1/4$-
and $3/4$-vison patterns with $2\times2$ plaquette unit cells.  We use
$J_0>0$ as the energy unit, so $t_{\rm DW}=2J_0$ at the loop point.  For
each pattern, we minimize the unprojected Majorana ground-state energy over
the four global Wilson-loop sectors without imposing the fermion-parity
projection.

Fig.~\ref{fig:sectorcuts}(a) shows the energy densities along $J_1=0$. At the
loop point the $2/3$-vison configuration reproduces the independent-hexagon
value $E_{\rm M}^{(0)}/N_s=-2t_{\rm DW}/3=-4J_0/3$, and at the pure Kitaev
point the zero-flux sector gives $-0.7872987J_0$ on the $120\times120$ torus,
against the thermodynamic $-0.7872986J_0$ from the exact zero-flux
dispersion~\cite{Kitaev2006}. Away from these limits the $2/3$-vison state
remains lowest over a finite interval around the loop point, the zero-flux
state is recovered toward the Kitaev limit, and a narrow $1/3$-vison window
separates them; linear interpolation places the crossings near
$J_2/J_0\simeq-0.0778$ and $-0.0135$. 
As shown in Fig.~\ref{fig:sectorcuts}(b), along the vertical cut $J_2/J_0=-1$,
the $2/3$-vison state stays lowest of the seven
throughout, with the stripe as the closest competitor; the separation falls
from $0.060\,J_0$ to $2.39\times10^{-2}J_0$ per site over the scanned
interval, less than a factor of three.

Two limitations should be kept distinct. At the loop point the selection does
not rely on the candidate set at all: Eq.~\eqref{eq:globalbound} bounds the
unprojected matter energy over all $2^{N_p-1}$ allowed local flux
configurations and all four holonomy sectors,
with the six additional patterns included only as special cases.
Away from the loop point, however, the comparison is restricted
to the candidate patterns in Table~\ref{tab:flux_patterns}.
The scan therefore establishes a robust energetic preference  for the $2/3$-vison crystal within this seven-pattern set,
but cannot exclude a lower-energy periodic configuration that was not sampled.
Establishing thermodynamic stability requires broader flux searches,
physical-state projection, and finite-size scaling~\cite{ZhangPRL2019,Nasu2014}.

\section{Discussion and outlook}
\label{sec:discussion}

Our central result is an exact, bidirectional feedback between gauge flux and
matter connectivity.   
At the loop point the bond term and its two-plaquette dressing cancel on
every equal-flux bond, so only flux-domain-wall bonds remain active.
Domain-wall parity at each
trivalent vertex restricts the active degree to zero or two, so every flux
sector decomposes into nonbranching Majorana rings and isolated sites.  The exact
energy decomposition of Sec.~\ref{subsec:ising_decomposition} closes the
feedback loop: a triangular-lattice Ising term selects the extensively
degenerate fully packed manifold, and a nonnegative Majorana residual then
selects, on admissible commensurate tori, precisely the three translated
$2/3$-vison crystals.  The bound applies to every local flux pattern and all
four Wilson sectors, rather than to a chosen set of candidates.   
This exact lifting is a nonperturbative
form of fermion-mediated order-by-disorder.  Earlier extended Kitaev
models~\cite{ZhangPRL2019,ZhangPRR2020} instead modulate third-neighbor hopping
without fragmenting the nearest-neighbor graph, and identify flux order by
comparing candidate sectors rather than through a global bound.

Gauge projection is essential for converting this matter-energy minimum into
a statement about the spin Hamiltonian.  The exact parity rule
\eqref{eq:parityrule_main}, which is independent of the Wilson sector, shows
that for $J_0<0$ the crystal vacuum is physical for every $N_{\rm hex}$.
Hence the unprojected all-sector bound proves that the crystal is a rigorous
spin ground state on admissible commensurate tori.  For $J_0>0$, the vacuum
survives only when $N_{\rm hex}$ is even.  If $N_{\rm hex}$ is odd, projection
requires one matter quantum and raises the crystal-sector energy by
$2t_{\rm DW}$.  This $O(1)$ correction leaves the thermodynamic energy density
unchanged but can reorder flux sectors at finite size.  
Projection therefore fixes both the domain of the exact
ground-state result and the finite-size level structure.

The matching point of Sec.~\ref{subsec:matching} provides the complementary limit.  
There the same two operators cancel with the opposite relative sign, deleting
domain-wall hoppings and turning triangular-Ising ground states into perfect
matchings, whereas the uniform flux-free sector activates the full honeycomb
graph, saturates the Lieb-based global bound on the unprojected matter energy,
and lies below every matching sector.  
At the loop point the connectivity rule is reversed:
only domain-wall bonds remain, so uniform flux eliminates hopping altogether.
Flux uniformity and kinetic connectivity therefore cannot be optimized
simultaneously.

Local unprojected flux rearrangements further expose this graph dependence.
Flipping an $A$ plaquette destroys an elementary ring and yields six isolated
Majorana zero modes, whereas flipping a $B$ or $C$ plaquette reconnects three
hexagons into one antiperiodic $L=18$ ring, with no zero mode and only about
one-eighth the energy cost.  These single-plaquette costs are not physical flux
gaps on a closed torus, where $\prod_p w_p=1$ restricts the allowed flux
changes.  They nevertheless show that zero modes diagnose connectivity and
twist, not an isolated vison.

Topological order requires a separate argument.  Fourfold Wilson-sector
degeneracy alone is inconclusive because contractible Majorana rings are automatically
holonomy independent.  Likewise, topological entanglement entropy $\gamma=\ln2$ for the ring-aligned
Kitaev--Preskill partition~\cite{Preskill2006} is supplied by the gauge
constraint, while the factorized matter rings contribute no topological term
(see Appendix~\ref{app:tee}).  The decisive result is instead the explicit
ground-space mapping of Sec.~\ref{sec:toriccode}.  Within the product of
six-site ring ground spaces, a depth-one local unitary and edge relabeling map
the physical $B$- and $C$-plaquette flux operators to the weight-four
$Z$-plaquette and $X$-star stabilizers of a sheared square-lattice toric code.
The compatible ground spaces---though not the full spectra---are therefore
exactly equivalent.  Each crystal branch carries two logical qubits, exact
fourfold degeneracy, and an effective code distance growing linearly with
system size at fixed aspect ratio. 
For $J_0<0$, and for $J_0>0$ with an even number of rings, each vison crystal branch 
thus realizes a toric-code ground space dressed by a depth-one product of six-spin unitaries; 
the three branches encode broken lattice-translation symmetry. 

We call the loop-point decomposition \emph{flux-conditioned graph
fragmentation}.  It is neither disorder-induced
Anderson localization~\cite{Anderson1958,Abrahams1979,Lee1985} nor many-body Hilbert-space
fragmentation~\cite{Sala2020,Khemani2020,Moudgalya2022}: each fixed-flux block remains a
quadratic Majorana problem.  The mechanism is closer to Aharonov--Bohm
caging~\cite{Vidal1998}, but with a key distinction---the conserved flux
configuration programs the kinetic graph itself rather than localizing motion
on a fixed graph.

The exact selection mechanism is fine-tuned to the loop point.  Bond anisotropy
that preserves the exact flux-conditioned bond cancellation retains the nonbranching graph but
makes ring energies depend on the bond sequence, invalidating the isotropic
hexagon bound.  Adiabatic continuity suggests stability to weak deviations as
long as the matter and physical flux gaps remain open, but this does not locate
a phase boundary.  Away from the loop point, the finite-torus comparison of
seven periodic patterns neither excludes unsampled flux orders nor establishes
the thermodynamic phase diagram; doing so requires broader flux searches.  
A quantitative domain-wall counting bound and the associated thermal melting
problem remain open.
 
The engineered multispin terms define an exactly solvable
parent Hamiltonian rather than a minimal microscopic model for candidate Kitaev materials~\cite{Jackeli2009,Trebst2022-go,Takagi2019-ok,Matsuda2025-pd,Rousochatzakis2024-td}, although related couplings may be accessible in programmable 
quantum simulators~\cite{Duan2003,Katz2023,Park2025,Will2025,Evered2025}.
In summary, the model realizes gauge-matter feedback explicitly: conserved
flux determines the geometry of Majorana-fermion hopping, while the resulting
matter energetics feeds back to select a vison crystal combining broken
translation symmetry with intrinsic $\Ztwo$ topological order.

\begin{acknowledgments}
J.W. acknowledges support from the National Natural Science Foundation of China (Grant No.~12404170) and the start-up grant at HZNU.
C.C. acknowledges support from the National Natural Science Foundation of China (Grants No.~12404175 and No.~12247101), the Fundamental Research Funds for the Central Universities (Grant No.~lzujbky-2025-jdzx07), and the Natural Science Foundation of Gansu Province (Grants No.~22JR5RA389 and No.~25JRRA799).
\end{acknowledgments}

\appendix

\section{Local sign identities and the hexagon twist}
\label{app:spinstrings}
\label{app:hexsign}

This appendix fixes the signs entering both the multispin Hamiltonian and the
elementary-ring twist.  Label a hexagon $1,\ldots,6$ as in
Fig.~\ref{fig:gaugestring}(a), with
\begin{equation}
W_p=\sigma_1^x\sigma_2^y\sigma_3^z
    \sigma_4^x\sigma_5^y\sigma_6^z .
\end{equation}
For $e=(12)_z$, Eq.~\eqref{eq:Spath} applied to the complementary five-bond
path in Fig.~\ref{fig:gaugestring}(b), together with
$\sigma^z\sigma^x=\ii\sigma^y$ and
$\sigma^z\sigma^y=-\ii\sigma^x$, gives
\begin{align}
\calS_{\bar e}
&=\ii^{\,4}B_{16}^yB_{65}^xB_{54}^zB_{43}^yB_{32}^x \nonumber\\
&=\sigma_1^y\sigma_2^x\sigma_3^z
  \sigma_4^x\sigma_5^y\sigma_6^z
 =B_{12}^zW_p .
\end{align}
Thus $B_eW_p$ is exactly the six-spin string on the complementary path.

For adjacent plaquettes $p$ and $q$ sharing $e=(12)_z$, the convention of
Eq.~\eqref{eq:Wp_spin} gives
$W_q=\sigma_1^y\sigma_2^x\mathcal O_4^{(q)}$, where $\mathcal O_4^{(q)}$
contains its four remaining Pauli factors.  At the shared endpoints,
\begin{align}
(\sigma_1^x\sigma_1^y)(\sigma_2^y\sigma_2^x)
&=(\ii\sigma_1^z)(-\ii\sigma_2^z)=B_e,\nonumber\\
B_eW_pW_q&=\mathcal O_8,
\end{align}
where
$\mathcal O_8=\sigma_3^z\sigma_4^x\sigma_5^y\sigma_6^z\mathcal O_4^{(q)}$
is supported on the eight outer sites of the two-plaquette cluster [see Fig.~\ref{fig:gaugestring}(c)].  
The endpoint phases therefore cancel, and the eight-spin term carries no extra sign.  
Unlike $B_eW_p$, the operator $B_eW_pW_q$ is not a spin string on
a single complementary path; it is the bond term multiplied by the two
plaquette-flux operators $W_pW_q$.
The combined relabeling $n\mapsto n+1$ and color cycle
$x\mapsto y\mapsto z\mapsto x$ extends these identities to every bond.
Together with $B_e=-\ii u_ec_ic_j$, they yield
Eqs.~\eqref{eq:majmodel} and~\eqref{eq:Jeff} without additional signs.

The same conventions determine the twist of an elementary Majorana ring.
The ordered product of its six bond operators is
\begin{equation}
B_{12}^zB_{23}^xB_{34}^yB_{45}^zB_{56}^xB_{61}^y=W_p .
\end{equation}
Let $u_e^{\rm trav}$ be oriented along the boundary traversal.  Using
$B_e=-\ii u_e^{\rm trav}c_ic_j$ gives
\begin{align}
W_p
&=(-\ii)^6
  \left(\prod_{e\in\partial p}^{\rm trav}u_e\right)
  (c_1c_2)\cdots(c_6c_1)\nonumber\\
&=-\prod_{e\in\partial p}^{\rm trav}u_e
=\prod_{e\in\partial p}^{A\to B}u_e .
\end{align}
Here the Majorana chain contracts to unity, and the last equality follows
because three links are traversed opposite to the fixed $A\to B$ orientation.
This also verifies Eq.~\eqref{eq:Wp_majorana}.  
Rewriting Eq.~\eqref{eq:HDW} in the loop traversal order and comparing with
Eq.~\eqref{eq:ringH} gives $s_n=\widetilde u_n^{\rm trav}=\operatorname{sgn}(J_0)u_n^{\rm trav}$; 
reversing a bond changes the signs of both $u_e$ and $c_ic_j$, leaving their product invariant.  Hence,
\begin{align}
\varphi_{\rm hex}
&=\prod_{n=1}^{6}s_n
=\bigl[\operatorname{sgn}(J_0)\bigr]^6
  \prod_{e\in\partial p}^{\rm trav}u_e
=-w_p .
\end{align}
Therefore, Eq.~\eqref{eq:loop_twist_general} holds for $L=6$ independently of the
sign of $J_0$.

\section{Ring-energy sums and lower bound}
\label{app:ring}
 
For an even-length Majorana ring, the unprojected vacuum energy is
\begin{equation}
E_0=-\frac12\sum_{\epsilon_a>0}\epsilon_a.
\end{equation}
For $\varphi=+1$, the positive single-particle energies correspond to
$k_m=2\pi m/L$, with $m=1,\ldots,L/2-1$.  Using the standard sine-sum
identity,
\begin{equation}
\sum_{m=1}^{L/2-1}\sin\frac{2\pi m}{L}
=
\cot\frac{\pi}{L},
\end{equation}
one obtains
\begin{equation}
E_0(L,+)
=
-2t_{\rm DW}\cot\frac{\pi}{L}.
\end{equation} 
For $\varphi=-1$, the allowed positive-energy momenta are
$k_m=(2m+1)\pi/L$, with $m=0,\ldots,L/2-1$, and
\begin{equation}
\sum_{m=0}^{L/2-1}
\sin\frac{(2m+1)\pi}{L}
=
\csc\frac{\pi}{L}.
\end{equation}
Hence one obtains
\begin{equation}
E_0(L,-)
=
-2t_{\rm DW}\csc\frac{\pi}{L}.
\end{equation}
These expressions give Eq.~\eqref{eq:ringenergy}, and $\cot x<\csc x$ on
$(0,\pi/2)$ gives $E_0(L,-)<E_0(L,+)$ at fixed $L$.
For the honeycomb loops considered here, $L\ge6$, so
$x=\pi/L\in(0,\pi/6]$.  The concavity of $\sin x$ gives
\begin{equation}
\sin x
\ge
\frac{\sin(\pi/6)}{\pi/6}x
=
\frac{3}{\pi}x,
\end{equation}
and therefore $\sin\frac{\pi}{L}\ge\frac{3}{L}$.
Using the antiperiodic branch,
\begin{equation}
E_0(L,-)
\ge
-\frac{2t_{\rm DW}}{3}L ,
\end{equation}
and since the periodic branch lies higher at the same $L$, the bound holds
for either twist.  The concavity bound is saturated only at $x=\pi/6$, and the
periodic branch is strictly higher, so equality requires
$(L,\varphi)=(6,-1)$.

\section{Spin ED and finite-torus parity}
\label{app:finite-torus-ed}

The loop bound selects the $2/3$-vison crystal at the level of the
unprojected Majorana vacuum energy.  On a finite torus, gauge projection may
reject that vacuum and reorder flux sectors.  We thus distinguish the
lowest physical energy of the crystal sector from the ground-state energy of
the full spin Hamiltonian.
The commuting plaquette fluxes $\{W_p\}$ and Wilson loops
$(\mathcal W_1,\mathcal W_2)$ decompose the physical spin Hilbert space into
flux--Wilson blocks; spin ED within such a block requires no further
projection.  In the Majorana formulation, by contrast, projection retains
only one total matter-parity sector in each block.  Its expression contains
convention-dependent gauge and ordering factors, although the resulting
physical parity and projected spectrum are gauge invariant and in general
depend on the Wilson sector~\cite{Pedrocchi2011,ZschockeVojta2015}.  In the
crystal sector, the required parity is instead identical in all four Wilson
sectors, as established below; this does not follow from holonomy independence
of the contractible-ring spectrum alone.  

At the loop point, we diagonalize the spin Hamiltonian $H_{\rm DW}^{\rm spin}$ in Eq.~\eqref{eq:HDWspin}.
A torus with $N_p$ unit cells contains $N_s=2N_p$ spins.  Because
$\prod_pW_p=1$, the $N_p-1$ independent plaquette fluxes together with the two
Wilson loops provide $N_p+1$ independent commuting operators, so each
flux--Wilson block has dimension $D=2^{N_p-1}$.
Here $\mathcal W_\mu$ is the ordered product of Kitaev bond operators
along a noncontractible cycle $\gamma_\mu$, with eigenvalue the link holonomy
$\mathcal W_{\gamma_\mu}$ of Eq.~\eqref{eq:global_holonomy}; this fixes the
correspondence between spin and Majorana sectors.
For $N_p=9$ and $12$ we built an orthonormal basis for the image of each
stabilizer projector, checked that its rank is $2^{N_p-1}$, and diagonalized the
block densely.  For the larger tori, we wrote the
stabilizers as binary symplectic vectors and used a tableau algorithm to find a
Clifford circuit mapping them to single-qubit $Z$ operators.  Conjugation leaves
the Hamiltonian a sum of Pauli strings; fixing the $N_p+1$ stabilized qubits
gives a Hamiltonian on $N_p-1$ logical qubits, solved by matrix-free Lanczos,
each string acting through a bit-flip pattern and a phase. 
In sectors without Majorana zero modes we identify the physical matter
parity by comparing the spin-block ground-state energy with the parity-resolved
free-Majorana values.  These differ by the sector's smallest positive
single-particle energy $\epsilon_{\min}$, so the assignment is unambiguous; the
agreement is better than $10^{-12}\,t_{\rm DW}$.   
On the $N_p=9$ torus we also diagonalized the full $2^{18}$-dimensional
spin space, without blocking, as an independent check.

Table~\ref{tab:app-ed-parity} summarizes the spin ED results for six consecutive
three-color-compatible tori.  Here
$N_{\rm hex}=N_p/3$ is the number of antiperiodic hexagonal Majorana
rings.  The unprojected 2/3-vison crystal energy and its lowest positive
single-particle energy are
\begin{equation}
 E_{\rm M}^{(0)}=-4t_{\rm DW}N_{\rm hex},\qquad
 \epsilon_{\min}=2t_{\rm DW}.
 \label{eq:app-ed-majorana-energy}
\end{equation}
To define the parity quoted below, we write a many-body eigenstate as
an occupation pattern of the positive-energy matter modes and set $N_f=\sum_a f_a^\dagger f_a$ and $P_f=(-1)^{N_f}$.
The unprojected quasiparticle vacuum has $N_f=0$ and is called even.
The parity column specifies which of the even and odd matter-Fock
subspaces survives projection onto the physical spin sector.  
It does \emph{not} denote either Wilson-loop eigenvalue or the parity of the vison number.  
The selected matter parity is identical in all four Wilson sectors for every torus studied.

The required parity can also be obtained directly from the
Majorana algebra.  Let $A_u$ denote the nonsingular crystal-sector matrix
in Eq.~\eqref{MajH}, and choose $\bm c=Q_u\bm\gamma$ with
$Q_u\in O(N_s)$ such that
\begin{equation}
 Q_u^T A_u Q_u=\bigoplus_{m=1}^{N_s/2}
 \begin{pmatrix}0&\epsilon_m\\-\epsilon_m&0\end{pmatrix},
 \qquad \epsilon_m>0.
 \label{eq:app-ed-canonical-A}
\end{equation}
Define $\mathcal U_u=\prod_{\langle ij\rangle_\alpha}^{A\to B}u_{ij}$ from the
original link variables $u_{ij}=\ii b_i^\alpha b_j^\alpha$, including
inactive bonds.  The active matter hopping contains
$\widetilde u_e=s u_e$, whereas $\mathcal U_u$ in the projection identity
contains the original $u_e$.  Let $\epsilon_{\rm perm}$ be the sign of the
permutation that rearranges
$\prod_i(b_i^xb_i^yb_i^zc_i)$ into link-paired $b$ Majoranas followed by
the ordered word $c_0\cdots c_{N_s-1}$.  Using
$b_i^\alpha b_j^\alpha=-\ii u_{ij}$ and
\begin{equation}
 \begin{aligned}
 c_0c_1c_2\cdots c_{N_s-1}
 &=\det(Q_u)\,\ii^{N_s/2}\widehat P_f,\\
 \det(Q_u)&=\sgn\!\operatorname{Pf}(A_u),
 \end{aligned}
 \label{eq:app-ed-pfaffian-relations}
\end{equation}
where
$f_m=(\gamma_{2m-1}+\ii\gamma_{2m})/2$ and
$\widehat P_f=(-1)^{\sum_m f_m^\dagger f_m}$, gives the exact operator
identity
\begin{equation}
 \prod_iD_i=\epsilon_{\rm perm}(-1)^{N_s/2}\mathcal U_u
 \sgn\!\operatorname{Pf}(A_u)\widehat P_f.
 \label{eq:app-ed-parity-identity}
\end{equation}
The above equation is restricted to nonsingular
$A_u$; sectors with zero modes require a separate choice of the zero-mode
Fock space before projection.

Although $\epsilon_{\rm perm}$, $\mathcal U_u$, and
$\sgn\operatorname{Pf}(A_u)$ separately depend on site ordering, link
orientation, and gauge, their product is invariant.
The value of that product in a crystal sector can be fixed without an
ordering-specific Pfaffian calculation.  Section~\ref{sec:toriccode}
constructs the eigenspace of $H_{\rm DW}^{\rm spin}$ at the $2/3$-vison crystal energy
directly: by the stabilizer ranks and the two redundancies
$\prod_p\mathcal B_p=\prod_{c'}\mathcal A_{c'}=\mathbb I$, that eigenspace is
four dimensional when $(-s)^{N_{\rm hex}}=1$ and empty otherwise.  The crystal
matter matrix is a direct sum of nonsingular antiperiodic hexagon blocks in
four Wilson sectors, so each sector contributes at most one state.
States from different Wilson sectors lie in different superselection sectors,
hence are orthogonal, and together they exhaust that eigenspace; dimension four
therefore forces exactly one surviving vacuum per sector in the compatible
case.  In the incompatible case all four vacua are rejected, and since
projection retains one global matter parity in each sector, that parity is odd.
This establishes $P_f^{\rm phys}=(-s)^{N_{\rm hex}}$ throughout the crystal
family, and with $N_s/2=3N_{\rm hex}$ substitution into
Eq.~\eqref{eq:app-ed-parity-identity} gives
\begin{equation}
 \epsilon_{\rm perm}\mathcal U_u
 \sgn\operatorname{Pf}(A_u)= s^{N_{\rm hex}}.
 \label{eq:app-ed-crystal-invariant}
\end{equation}
For $J_0>0$ this product is unity.  Equation~\eqref{eq:app-ed-crystal-invariant}
thus follows from the spin constraint algebra and the exact projection
identity rather than from an inversion count.  Its holonomy independence
involves both the link product and the Pfaffian sign; holonomy independence of
the matter eigenvalues alone would not establish it.

\begin{table}[t]
 \centering
 \renewcommand{\arraystretch}{1.12}
 \setlength{\tabcolsep}{2pt}
 \resizebox{\columnwidth}{!}{
 \begin{tabular}{c c c c c c c c c c}
  \hline\hline
  & & & & & & \multicolumn{2}{c}{$J_0>0$}
  & \multicolumn{2}{c}{$J_0<0$} \\
  $N_s$ & $N_p$ & $N_{\rm hex}$ & $\mathbf T_1$ & $\mathbf T_2$
  & $E^{(0)}_{\rm M}$ & $P_f^{\rm phys}$ & $E_{\rm spin}^{\rm cr}$
  & $P_f^{\rm phys}$ & $E_{\rm spin}^{\rm cr}$\\
  \hline
  $18$  & $9$  & $3$ & $(3,0)$  & $(0,3)$  & $-12$ & odd   & $-10$ & even & $-12$ \\
  $24$  & $12$ & $4$ & $(4,-2)$ & $(0,3)$  & $-16$ & even  & $-16$ & even & $-16$ \\
  $30$  & $15$ & $5$ & $(5,-1)$ & $(0,3)$  & $-20$ & odd   & $-18$ & even & $-20$ \\
  $36$  & $18$ & $6$ & $(5,-1)$ & $(2,-4)$ & $-24$ & even  & $-24$ & even & $-24$ \\
  $42$  & $21$ & $7$ & $(4,1)$  & $(5,-4)$ & $-28$ & odd   & $-26$ & even & $-28$ \\
  $48$  & $24$ & $8$ & $(6,0)$  & $(2,-4)$ & $-32$ & even  & $-32$ & even & $-32$ \\
  \hline\hline
 \end{tabular}
 }
\caption{Direct spin ED at the loop point for both signs of $J_0$.  Translation
vectors are in primitive-cell coordinates; energies are in units of
$t_{\rm DW}$.  Here $E_{\rm spin}^{\rm cr}$ is computed from the spin
Hamiltonian within the specified $2/3$-vison crystal sector, and equals the
global ground-state energy only where Eq.~\eqref{eq:globalbound} is saturated.
The tori with $N_s=18$, $24$, and $30$ have systole six and are not admissible
in the sense of Sec.~\ref{sec:loops}; admissibility enters only the uniqueness
statement of Sec.~\ref{subsec:uniqueness}, whereas the parity rule tested here
concerns a given flux configuration and holds on any commensurate torus; including them therefore widens the test rather than weakening it.
}\label{tab:app-ed-parity}
\end{table}

Denote the surviving physical matter parity by $P_f^{\rm phys}$.  
Eqs.~\eqref{eq:app-ed-parity-identity} and
\eqref{eq:app-ed-crystal-invariant} give the exact crystal-sector rule
\begin{equation}
 \begin{aligned}
 P_f^{\rm phys}&=(-s)^{N_{\rm hex}},\\
 E_{\rm spin}^{\rm cr}-E_{\rm M}^{(0)}
 &=t_{\rm DW}\bigl[1-(-s)^{N_{\rm hex}}\bigr].
 \label{eq:app-ed-parity-rule}
 \end{aligned}
\end{equation}
For $J_0>0$ ($s=+1$), this becomes
$P_f^{\rm phys}=(-1)^{N_{\rm hex}}$: an odd number of rings removes the
even vacuum and costs one excitation, $2t_{\rm DW}$.  For $J_0<0$
($s=-1$), it becomes $P_f^{\rm phys}=+1$ for every $N_{\rm hex}$, so the
vacuum is always physical and the projection cost vanishes.  This is an
even--odd effect in $N_{\rm hex}=N_p/3=N_s/6$, not in the necessarily even
number $N_s$ of spins.  
The rule applies to commensurate crystal tori whose active components are
contractible elementary hexagons; it does not cover arbitrary flux sectors,
zero-mode sectors, or tori with noncontractible length-six components.

Whenever the even vacuum is physical, the crystal reaches the absolute
loop bound $E_{\rm spin}^{\rm cr}=-(2/3)t_{\rm DW}N_s$.  This occurs for even
$N_{\rm hex}$ at $J_0>0$ and, notably, for every $N_{\rm hex}$ at
$J_0<0$.  Since no projected state can lie below the unprojected all-sector
bound, on the $J_0<0$ branch the crystal is a rigorous ground state
of the spin Hamiltonian at every commensurate size, with no residual
variational uncertainty.  Its three translations and four Wilson
sectors give exactly $12$ states at the bound: the saturating class consists of
contractible elementary hexagons alone, since the honeycomb girth is six and
noncontractible length-six cycles are excluded by the systole condition of Sec.~\ref{sec:loops}.  
The three smallest tori listed in Table~\ref{tab:app-ed-parity},
$N_s=18$, $24$, and $30$, have systole six and therefore are outside of this condition.
The conclusion nevertheless holds there.  Disjoint noncontractible loops on
a torus are parallel and so share one nonzero $\Ztwo$ homology class; since the
active edge set bounds the positive-flux region, it is $\Ztwo$-null homologous
and such loops occur only in even numbers.  A covering built entirely from
noncontractible six-cycles needs $N_s/6$ of them, odd for $N_s=18$ and $30$,
and is thereby excluded.  The argument leaves mixed coverings open, and does
not apply at $N_s=24$, where two local-flux configurations do give four
noncontractible six-cycles---two periodic and two antiperiodic, hence
$2E_0(6,+)+2E_0(6,-)=-8-4\sqrt3$ rather than $-16$ in units of $t_{\rm DW}$.
Exhaustive enumeration settles all three: over the $4\times2^{N_p-1}$
flux--Wilson sectors the minimum is attained by exactly three local-flux
configurations, the three crystal translations, each in all four Wilson
sectors, whose twelve Majorana vacua all survive projection on 
the $J_0<0$ branch.
The resulting twelvefold degeneracy is exact rather than exponentially
split: the three translated patterns saturate the same bound, and each one is a direct sum of contractible six-site blocks whose
spectrum is strictly holonomy independent.

Only on the $J_0>0$ branch with odd $N_{\rm hex}$ does the
physical crystal contain one matter excitation and rise by $2t_{\rm DW}$.  On
the $3\times3$ torus, we additionally searched all $2^8\times4$ flux and Wilson sectors. 
The physical global minimum is a single $L=18$ loop with
\begin{equation}
 E_{\rm M}^{L=18}=-2t_{\rm DW}\csc\frac{\pi}{18}
 +4t_{\rm DW}\sin\frac{\pi}{18}\approx-10.8229\,t_{\rm DW},
 \label{eq:app-ed-long-ring}
\end{equation}
below $E^{\rm cr}_{\rm spin}=-10\,t_{\rm DW}$.  Full ED in the
$2^{18}$-dimensional spin basis confirms this without reference to flux
sectors, returning $E_0/t_{\rm DW}=-12$ for $J_0<0$, equal to
$E^{\rm cr}_{\rm spin}$, and $-10.8229$ for $J_0>0$, matching
Eq.~\eqref{eq:app-ed-long-ring}.  Projection thus reverses the ordering of
flux sectors on this torus.
 
The projection correction to the crystal energy,
\begin{equation}
 E_{\rm spin}^{\rm cr}-E_{\rm M}^{(0)}
 =\begin{cases}
  2t_{\rm DW}, & \quad J_0>0,\ N_{\rm hex}\ \text{odd},\\
  0, & \quad \text{otherwise},
 \end{cases}
\end{equation}
is exact at finite size.  It is $O(1)$, so both branches share the same
thermodynamic energy density, the correction to which scales as $N_s^{-1}$.

\section{Consistency check for the TEE}
\label{app:tee}
 
We first evaluate the Kitaev--Preskill topological entanglement entropy (TEE) for one
translation-breaking branch and one Wilson-sector ground state of the
$2/3$-vison crystal at the loop point.  We restrict to $J_0<0$, or $J_0>0$ with
even $N_{\rm hex}$, so that the projected Majorana vacuum is nonzero.  
For the translated flux pattern $a$ and Wilson sector $\sigma$, the state is
\begin{equation}
 |\Psi_{a,\sigma}\rangle \propto P_{\rm phys}
 \Big[\,|u^{(a)}_{\rm crystal}\rangle\otimes
 \bigotimes_{r=1}^{N_{\rm hex}}|{\rm g.s.}(6,-)\rangle_r\Big],
 \label{eq:app-tee-state}
\end{equation}
where $P_{\rm phys}=\prod_i(1+D_i)/2$.  Before projection the matter state is a
product over vertex-disjoint six-site rings.  We use the gauge--matter entropy
decomposition of Ref.~\cite{Yao2010} together with the Kitaev--Preskill
construction~\cite{Preskill2006,Levin2006}.

For a contractible region $R$ with one boundary component, let $N_{\partial R}$
denote the number of cut honeycomb bonds.  The gauge--matter decomposition
gives
\begin{equation}
 S(R)=S_G(R)+S_M(R),\quad
 S_G(R)=\left(\frac{N_{\partial R}}2-1\right)\ln2,\nonumber
\end{equation}
with $S_M$ the Gaussian matter entropy: the cut-bond gauge Majoranas supply the
boundary term, and a single boundary constraint reduces the entropy by
$\ln2$~\cite{Yao2010}.  That constraint is the ordered operator identity
\begin{align}
 1=\prod_{i\in R}D_i
 =\epsilon_R(-\ii)^{E_{\rm int}}
 \Big(\prod_{e\subset R}u_e\Big)
 \Big(\prod_{k=1}^{N_{\partial R}}\beta_k\Big)
 \Big(\prod_{i\in R}c_i\Big),\nonumber
\end{align}
where $E_{\rm int}$ counts internal bonds, $\beta_k$ are the gauge Majoranas
left unpaired by the cut, and $\epsilon_R=\pm1$ fixes the site, bond, and
Majorana ordering conventions; each internal contraction supplies
$b_i^\alpha b_j^\alpha=-\ii u_{ij}$ in the chosen orientation.  The individual
Majorana words are not Hermitian, and the phase $(-\ii)^{E_{\rm int}}$ is what
renders the product real: for odd $E_{\rm int}$ the word
$\bigl(\prod_k\beta_k\bigr)\bigl(\prod_{i\in R}c_i\bigr)$ is antihermitian and
supplies the compensating factor.  
 
Choose the disjoint adjacent regions $A$, $B$, and $C$ to be unions of complete
rings.  No cut then divides a ring, and hence
\begin{equation}
 S_M(X)=0,\quad X\in\{A,B,C,AB,AC,BC,ABC\},
 \label{eq:app-tee-aligned}
\end{equation}
every region entering the combination being contractible with one boundary
component.  For
\begin{equation}
 I_3=S_A+S_B+S_C-S_{AB}-S_{AC}-S_{BC}+S_{ABC},
 \label{eq:app-tee-kp}
\end{equation}
the boundary contributions cancel, leaving
\begin{equation}
 I_3=-\ln2,\qquad \gamma\equiv-I_3=\ln2.
 \label{eq:app-tee-result}
\end{equation}
 
A general cut can give nonzero matter entropy for an individual region.  The
contribution of any one ring to $I_3$ nevertheless cancels unless that ring
meets all four of $A$, $B$, $C$, and the complement of $ABC$.  Such a ring would
have to reach from the junction of the three regions to the outer boundary, so
none exists once that separation exceeds one ring diameter; the rings here have
diameter two lattice constants.  Equation~\eqref{eq:app-tee-result} therefore
holds for nonaligned partitions in any macroscopic Kitaev--Preskill
arrangement.  
 
Finally, a superposition of translated crystal branches carries a
symmetry-breaking contribution.  If the weights are $p_a=|c_a|^2$, every region
large enough to distinguish the three flux patterns acquires the Shannon term
$H(p)=-\sum_ap_a\ln p_a$; since the coefficients in
Eq.~\eqref{eq:app-tee-kp} sum to $3-3+1=1$, the combination shifts by $+H(p)$
and
\begin{equation}
 \gamma_{\rm cat}=\ln2-H(p),
 \label{eq:app-tee-cat}
\end{equation}
reducing to $(\ln2-\ln3)$ for equal weights.  Working in one branch therefore
isolates the intrinsic $\ln2$ term, whose topological interpretation 
provides a consistency check of the depth-one local-unitary mapping in
Sec.~\ref{sec:toriccode}.

\section{Toric-code mapping and numerical checks}
\label{app:local-indistinguishability}
  
\subsection{The hexagon algebra and effective qubits per ring}
\label{app:hexalgebra}
 
Let $\mathcal A_r$ be the algebra generated by the six bond operators
$B_1,\ldots,B_6$ of $\partial p_r$ in Eq.~\eqref{eq:cluster-form}.
The $2^6$ ordered monomials
\begin{equation}
B(\bm s)\equiv B_1^{s_1}B_2^{s_2}\cdots B_6^{s_6},
\qquad s_n\in\{0,1\},
\end{equation}
are distinct Pauli strings and therefore form a basis of $\mathcal A_r$,
so $\dim\mathcal A_r=64$.
Such a monomial commutes with every $B_m$ if and only if
$s_{m-1}=s_{m+1}$ for all $m$, with indices understood modulo $6$, and thus the
solutions are $\pmb s=\emptyset,\{1,3,5\},\{2,4,6\}$ and $\{1,\ldots,6\}$.  The center
of $\mathcal A_r$ is therefore four dimensional,
\begin{equation}
 \mathcal Z_r=\operatorname{span}\bigl\{1,\;\Theta_r,\;\bar\Theta_r,\;W_{p_r}\bigr\},
 \quad
 \begin{aligned}
 \Theta_r&=B_1B_3B_5,\\
 \bar\Theta_r&=B_2B_4B_6,
 \end{aligned}
 \label{eq:app-center}
\end{equation}
with $\Theta_r\bar\Theta_r=W_{p_r}$.  
Hence
$\mathcal A_r$ has four irreducible blocks, and
$\sum_{i=1}^{4}d_i^2=\dim\mathcal A_r=64$ has the unique solution $d_i=4$,
so $\mathcal A_r\cong M_4(\mathbb C)^{\oplus4}$.  Because $\Theta_r$, $\bar\Theta_r$, and
$W_{p_r}$ are traceless, the four central sectors of the $64$-dimensional
six-spin space are $16$ dimensional: every irreducible block carries
multiplicity four.  
The above results are the standard irrep--multiplicity decomposition of a finite-dimensional
operator algebra, with the commutant acting on the multiplicity
factor~\cite{Knill2000,Zanardi2001}.
The ground
space $\mathcal V_r$ of $h_r$ is one such multiplicity space:
\begin{equation}
 \dim\mathcal V_r=4,
 \qquad
 h_r\big|_{\mathcal V_r}=-4t_{\rm DW}\,\mathbb I_{\mathcal V_r},
 \label{eq:app-Vr}
\end{equation}
which establishes the four-dimensional ground space used in
Appendix~\ref{app:finite-torus-ed} and reproduces $E_0(6,-)$ of Eq.~\eqref{eq:hexenergies}. 
In $\mathcal V_r$ the central elements are scalars with values
\begin{equation}
 \Theta_r=\bar\Theta_r=s\equiv\sgn(J_0),
 \qquad
 W_{p_r}=+1
 \quad\text{on }\mathcal V_r,
 \label{eq:app-central-values}
\end{equation}
fixed by two conserved chagres.  First, the four blocks split into $W_{p_r}=\pm1$
pairs, which correspond through $\varphi_{\rm hex}=-w_p$ to the antiperiodic
and periodic rings of Eq.~\eqref{eq:hexenergies}, with ground energies
$-4t_{\rm DW}$ and $-2\sqrt3\,t_{\rm DW}$: the ground block has $W_{p_r}=+1$
for either sign of $J_0$.  Second, conjugation by
$\prod_{n=1}^{6}\sigma_n^{\alpha_n}$, with $\alpha_n$ the color of bond
$(n,n{+}1)$, flips every $B_e$ exactly once, so $h_r\to-h_r$ while $\Theta_r$
and $\bar\Theta_r$ change sign at fixed $W_{p_r}$: the two $W_{p_r}=+1$ blocks
are exact spectral mirrors under $J_0\to-J_0$, and the block spectra of
Table~\ref{tab:app-hexagon-blocks} yield $\Theta_r=\bar\Theta_r=s$.  The flux-free condition
$w_A=+1$ of the crystal is thus not an assumption but an output of the ground space of the cluster.

\begin{table}[t]
\centering
\renewcommand{\arraystretch}{1.12}
\begin{tabular}{c c}
\hline\hline
$(\Theta_r,\bar\Theta_r)$ & \quad Eigenvalues of $h_r/t_{\rm DW}$\\
\hline
$(+1,+1)$ & \quad  $-4,\;0,\;2,\;2$\\
$(-1,-1)$ & \quad $-2,\;-2,\;0,\;4$\\
$(+1,-1)$ or $(-1,+1)$ & \quad $-2\sqrt3,\;0,\;0,\;2\sqrt3$\\
\hline\hline
\end{tabular}
\caption{Spectra within the four-dimensional irreducible blocks of one hexagon,
in units of $t_{\rm DW}$ at $s=+1$; changing $s$ reverses all listed energies.
Each eigenvalue occurs with multiplicity four in the corresponding
$16$-dimensional central sector.  The two blocks with
$W_{p_r}=\Theta_r\bar\Theta_r=-1$ share the periodic-ring spectrum, with ground
energy $-2\sqrt3\,t_{\rm DW}=E_0(6,+)$, whereas the lowest block
$(\Theta_r,\bar\Theta_r)=(s,s)$ is antiperiodic and nondegenerate within its
factor.}
\label{tab:app-hexagon-blocks}
\end{table}

The two effective qubits are carried by the commutant.  Each site of
$\partial p_r$ lies on one neighboring $B$ and one neighboring $C$ plaquette, and
consecutive sites share one of them [Fig.~\ref{fig:toricmap}(a)].  Writing
$\beta_r^{(a)}$ and
$\gamma_r^{(a)}$, $a=1,2,3$, for the two-spin factors that the neighboring $B$
and $C$ plaquette operators contribute inside hexagon $r$,
\begin{equation}
 \begin{aligned}
 \beta_r^{(1)}&=\sigma_1^y\sigma_2^x, &
 \beta_r^{(2)}&=\sigma_3^x\sigma_4^z, &
 \beta_r^{(3)}&=\sigma_5^z\sigma_6^y, \\
 \gamma_r^{(1)}&=\sigma_2^z\sigma_3^y, &
 \gamma_r^{(2)}&=\sigma_4^y\sigma_5^x, &
 \gamma_r^{(3)}&=\sigma_6^x\sigma_1^z.
 \end{aligned}
 \label{eq:app-betagamma}
\end{equation}
Each of these commutes with all six $B_e$ of the hexagon, hence lies in
$\mathcal A_r'$ and acts purely on the multiplicity space, and
\begin{equation}
 \beta_r^{(1)}\beta_r^{(2)}\beta_r^{(3)}=\bar\Theta_r,
 \quad
 \gamma_r^{(1)}\gamma_r^{(2)}\gamma_r^{(3)}=\Theta_r.
 \label{eq:app-betagamma-products}
\end{equation}
Within each family the operators commute, being supported on disjoint pairs of
sites, while $\beta_r^{(a)}$ anticommutes with exactly the two $\gamma_r^{(b)}$
sharing a site with it.  These are the relations of a two-qubit Pauli algebra:
choosing the eigenbasis of $\beta_r^{(1)},\beta_r^{(2)}$ and fixing the residual
phases,
\begin{equation}
 \begin{aligned}
 \beta_r^{(1)}&\mapsto Z_1, &
 \beta_r^{(2)}&\mapsto Z_2, &
 \beta_r^{(3)}&\mapsto sZ_1Z_2,\\
 \gamma_r^{(3)}&\mapsto X_1, &
 \gamma_r^{(2)}&\mapsto X_2, &
 \gamma_r^{(1)}&\mapsto sX_1X_2,
 \end{aligned}
 \label{eq:app-pauliframe}
\end{equation}
up to signs fixed by Eq.~\eqref{eq:app-central-values}.  Every hexagon therefore
supplies two effective qubits, on which the $B$-side operators act as $Z$-type
and the $C$-side operators as $X$-type.
 
Each $B$ or $C$ plaquette has its six sites distributed two-by-two over three
neighboring $A$ hexagons, and the two-site factor it contributes inside hexagon
$r$ is one of the operators of Eq.~\eqref{eq:app-betagamma}.  Thus we obtain
\begin{equation}
 W_{p\in B}=\prod_{k=1}^{3}\beta^{(a_k)}_{r_k},
 \qquad
 W_{p\in C}=\prod_{k=1}^{3}\gamma^{(a_k)}_{r_k}.
 \label{eq:app-BC-factorization}
\end{equation}
By Eq.~\eqref{eq:app-pauliframe}, each product is a pure $Z$-type or
$X$-type Pauli operator.  It is supported on three hexagons but has effective
Pauli weight four, because one of the three factors is $Z_1Z_2$ or $X_1X_2$,
which is made explicit in Appendix~\ref{app:toric-explicit}.
 
Because each of the three $\beta_r^{(a)}$ of a hexagon is used by exactly one
$B$ plaquette, and likewise for the $\gamma_r^{(a)}$, taking the product of
Eq.~\eqref{eq:app-BC-factorization} over all $B$ or all $C$ plaquettes collects
each factor once and gives Eq.~\eqref{eq:two-relations} via
Eq.~\eqref{eq:app-betagamma-products}.  Both products are therefore scalars on
$\bigotimes_r\mathcal V_r$.  A third relation would require a proper subset of
the stabilizers to multiply to a scalar; since the $B$-type and $C$-type
generators are respectively pure $Z$ and pure $X$ in the frame of
Eq.~\eqref{eq:app-pauliframe}, any such relation would have to hold within one
type alone.  There the constraint is local: on each hexagon the three
$\beta_r^{(a)}$ act as $Z_1$, $Z_2$, and $s Z_1Z_2$, and a scalar arises
only from none or all three; selecting one $B$ plaquette therefore forces all
others on a connected commensurate torus.  There are thus exactly two relations, as used in
Eq.~\eqref{eq:code-dim}.

\subsection{Explicit mapping to the square-lattice toric code}
\label{app:toric-explicit}
 
The counting of Sec.~\ref{sec:toriccode} gives $k=2$. We now make
the toric-code structure explicit by relabeling the effective
qubits as edges of a square lattice.
 
Label every crystal hexagon as in Fig.~\ref{fig:toricmap}(a), with site $1$ at the
top. For a hexagon centered at $\pmb v$, let $\pmb n_a$ ($a=1,2,3$) 
be the vectors from $\pmb v$ to the centers of its three neighboring $B$
plaquettes, across the bonds $(12)$, $(34)$, $(56)$, and let $\pmb m_a$ be
the corresponding vectors to the $C$ plaquettes across $(23)$, $(45)$,
$(61)$; explicitly, $\pmb n_1,\pmb n_2,\pmb n_3$ point at $60^{\circ}$,
$-60^{\circ}$, $180^{\circ}$ and $\pmb m_1,\pmb m_2,\pmb m_3$ at $0^{\circ}$,
$-120^{\circ}$, $120^{\circ}$, with $\sum_a \pmb n_a=\sum_a \pmb m_a=0$ and the
identities
\begin{equation}
\pmb m_3-\pmb m_1 = \pmb n_3-\pmb n_2,\quad \pmb m_3-\pmb m_2 = \pmb n_1-\pmb n_2.
\label{eq:direction-identities}
\end{equation}
Because all hexagons carry identical labels, the $B$ plaquette
centered at $b$ receives the factor $\beta^{(a)}$ from the hexagon at
$\pmb b-\pmb n_a$, and the $C$ plaquette at $\pmb c$ receives $\gamma^{(a)}$ from the
hexagon at $\pmb c-\pmb m_a$. In the frame of Eq.~\eqref{eq:app-pauliframe}, the effective stabilizers are therefore
\begin{equation}
\begin{aligned}
W_{p\in B}(\pmb b) &\longmapsto
  s\,Z_1(\pmb b-\pmb n_1)\,Z_2(\pmb b-\pmb n_2)\,\bigl[Z_1Z_2\bigr](\pmb b-\pmb n_3),\\
W_{p\in C}(\pmb c) &\longmapsto
  s\,\bigl[X_1X_2\bigr](\pmb c-\pmb m_1)\,X_2(\pmb c-\pmb m_2)\,X_1(\pmb c-\pmb m_3).
\end{aligned}
\label{eq:weight4-checks}
\end{equation}
Each stabilizer is supported on three hexagons but has Pauli weight
\emph{four} on the effective qubits.
 
Define the square-lattice axes
\begin{equation}
\hat{x}\equiv \pmb n_3-\pmb n_1,\qquad \hat{y}\equiv \pmb n_3-\pmb n_2,
\label{eq:square-axes}
\end{equation}
two nearest-neighbor vectors of the triangular lattice of $A$
hexagons, and identify hexagon $v$ with the square-lattice vertex $\pmb v$,
its effective qubit $2$ with the edge $e_x(\pmb v)=[\pmb v,\pmb v+\hat{x}]$, and its
effective qubit $1$ with the edge $e_y(\pmb v)=[\pmb v,\pmb v+\hat{y}]$. Rebasing
Eq.~\eqref{eq:weight4-checks} at $\pmb p\equiv \pmb b-\pmb n_3$ and $\pmb c'\equiv \pmb c-\pmb m_1$
and using Eq.~\eqref{eq:direction-identities} gives
\begin{equation}
\begin{aligned}
W_{p\in B}(\pmb b) &\longmapsto
  s\,Z_{e_x(\pmb p)}Z_{e_y(\pmb p)}Z_{e_y(\pmb p+\hat{x})}Z_{e_x(\pmb p+\hat{y})}
  \equiv s\,\mathcal{B}_{p},\\
W_{p\in C}(\pmb c) &\longmapsto
  s\,X_{e_x(\pmb c')}X_{e_y(\pmb c')}X_{e_x(\pmb c'-\hat{x})}X_{e_y(\pmb c'-\hat{y})}
  \equiv s\,\mathcal{A}_{c'}.
\end{aligned}
\label{eq:toric-checks}
\end{equation}
These are precisely the plaquette and star operators of the toric code on
the square lattice of hexagons, each multiplied by the scalar $s$.  No
additional Clifford transformation is required: the identification is only
a relabeling of qubits and lattice directions and therefore preserves the
depth-one character of local unitary $U$.  Equation~\eqref{eq:toric-checks} thus provides
the explicit coordinate realization of Eq.~\eqref{eq:toric-checks-main};
the projection compatibility, code dimension, and logical structure follow
as described in Sec.~\ref{sec:toriccode}.

\subsection{Numerical confirmation on the $N_p=9$ torus}
\label{app:ed-check}
 
We now verify the construction numerically by exact stabilizer-projected spin
ED. For a fixed crystal translation, let
$\mathcal G=\operatorname{span}\{\lvert\Psi_{w_1w_2}\rangle\}$ and denote its
projector by $P_{\mathcal G}$.  The topological-order condition is
\begin{equation}
 P_{\mathcal G}O_{\rm loc}P_{\mathcal G}=c_OP_{\mathcal G},
 \quad
 \langle\Psi_a\rvert O_{\rm loc}\lvert\Psi_b\rangle=c_O\delta_{ab}.
 \label{eq:app-local-indistinguishability}
\end{equation}
The scalar $c_O$ may depend on position relative to the three-sublattice
order, but not on $(w_1,w_2)$.
 
We evaluated Eq.~\eqref{eq:app-local-indistinguishability} on the $3\times3$
torus $\bm T_1=(3,0)$, $\bm T_2=(0,3)$, at $J_0=-1$, for which the Majorana
vacuum is physical and unique in every Wilson sector.  The four states have
$E_0=-12t_{\rm DW}=-24$, and the lowest excitation within each fixed flux and
Wilson sector lies at $\Delta_{\rm M}=4t_{\rm DW}=8$, confirming
Eq.~\eqref{eq:gap_phys}.

\begin{table}[t]
 \centering
 \renewcommand{\arraystretch}{1.12}
 \setlength{\tabcolsep}{3.3pt}
 \resizebox{\columnwidth}{!}{
 \begin{tabular}{l r c c}
  \hline\hline
  Operator set & Number & $\max_a|M_{aa}-\overline M|$
  & $\max_{a\ne b}|M_{ab}|$ \\
  \hline
  single-site $\sigma_i^\alpha$ & 54 & $0$ & $0$ \\
  Kitaev bonds $B_e^\alpha$ & 27 & $\lesssim 2.5\times10^{-15}$ & $0$ \\
  two-point, graph distance 1 & 243 & $\lesssim 2.5\times10^{-15}$ & $0$ \\
  two-point, graph distance 2 & 486 & $0$ & $0$ \\
  complete bases on all hexagons & 36855 & $\lesssim 2.5\times10^{-15}$ & $0$ \\
  \hline\hline
 \end{tabular}
 }
 \caption{Local-indistinguishability test on the $N_p=9$ torus at
 $J_0=-1$, with $M_{ab}=\langle\Psi_a\rvert O_{\rm loc}\lvert\Psi_b\rangle$ and
 $\overline M=\frac14\sum_aM_{aa}$.  The last row contains
 $9(4^6-1)$ nonidentity Pauli strings and, together with the identity,
 forms a complete operator basis on every elementary hexagon.}
 \label{tab:app-local-ind}
\end{table}

Table~\ref{tab:app-local-ind} shows that, within numerical precision, all
tested local operators act as scalars on $\mathcal G$.  
Consequently, the four ground states have
identical reduced density matrices on any such hexagon, while their
off-diagonal reduced transition operators vanish.  No hexagon-supported
operator can therefore distinguish or mix the four states, and any
perturbation that is a sum of such terms produces only a common first-order
energy shift.  The off-diagonal zeros follow exactly from stabilizer
selection rules; the residual diagonal spreads are at floating-point precision.

Note that the spin ED verifies rather than establishes this
mapping onto the square-lattice toric code, since logical strings and tested local operators
have comparable support on the smallest commensurate torus.

\begin{figure*}[t]
 \centering
 \includegraphics[width= \linewidth]{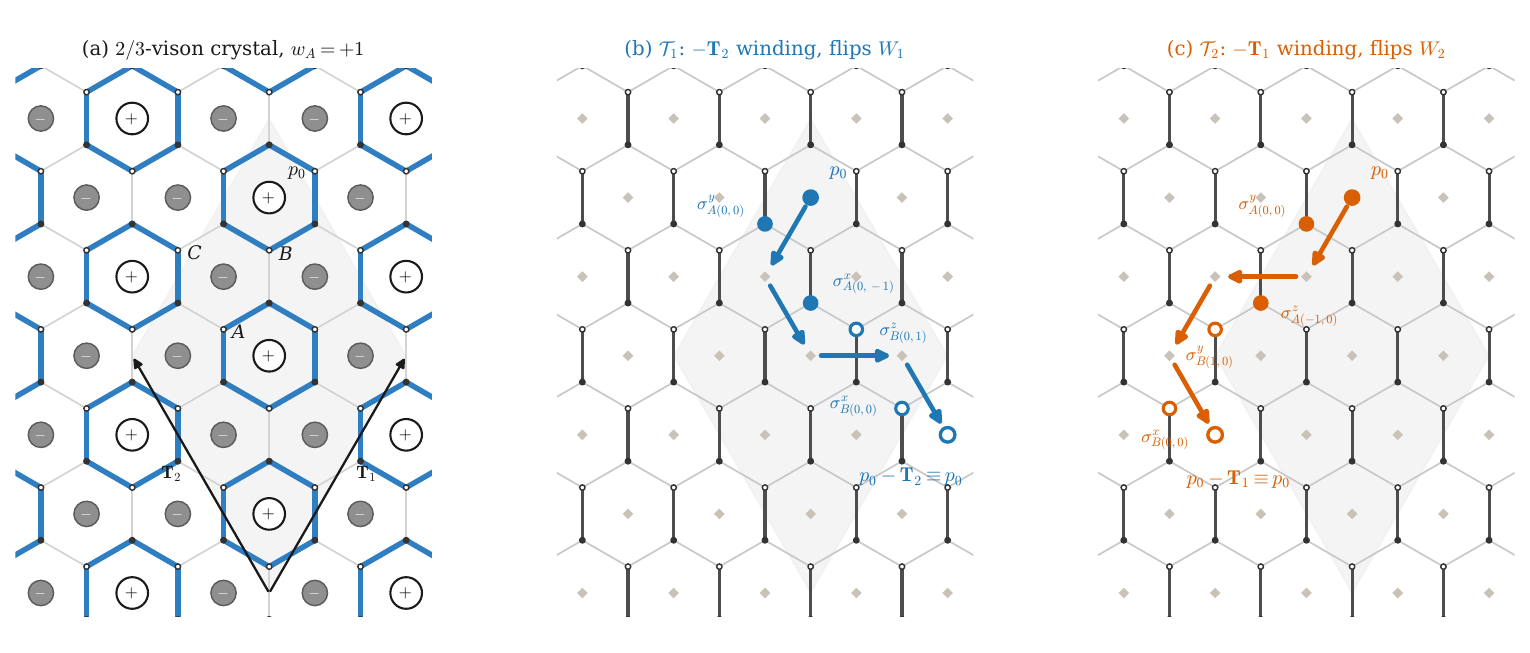}
 \caption{(a) The $2/3$-vison crystal on a $3\times3$ torus. Flux-free $A$ plaquettes (circled $+$) are surrounded by visons (filled circles, $w_p=-1$). Thick blue domain-wall bonds form $N_{\mathrm{hex}}=3$ vertex-disjoint antiperiodic hexagonal Majorana rings covering all $N_s=18$ sites. The shaded rhombus, spanned by $\bm T_1$ and $\bm T_2$, is a fundamental domain; the pattern repeats outside it.
(b),(c) Closed dual-lattice Pauli strings $\mathcal T_1$ and $\mathcal T_2$ [Eq.~\eqref{eq:app-dual-loops}], respectively, overlaid on the physical honeycomb lattice within the same domain. 
Dark vertical links are $z$ bonds; their filled lower and open upper endpoints denote $A$ and $B$ sites, respectively. 
Pale diamonds mark hexagon centers on the dual triangular lattice. The colored filled and open endpoints represent the same quotient plaquette; they differ by $-\bm T_2$ in (b) and $-\bm T_1$ in (c). Arrows follow the factor order in Eq.~\eqref{eq:app-dual-loops}; colored site markers identify the corresponding Pauli factors, with filled/open denoting the $A/B$ sublattices.
}\label{fig:app-dual-loops}
\end{figure*}

\subsection{Noncontractible dual-lattice strings}
\label{app:strings}
 
We finally test whether noncontractible dual strings act within, rather than
merely between sectors of, the four-dimensional ground space.  The test is
carried out on the $3\times3$ torus; of what follows, only the string weight
depends on that choice, growing as $O(k)$ on a $3k\times3k$ torus in parallel
with the code distance $d_{\rm eff}=\min(\ell_{\rm sys},\ell^{*}_{\rm sys})$ of
Sec.~\ref{sec:toriccode}.  We construct two weight-four closed
dual-lattice ('t Hooft) Pauli strings, using
physical honeycomb unit-cell coordinates with
\begin{equation}
 \begin{aligned}
  &\bm a_1=(\sqrt3/2,3/2),\quad
  \bm a_2=(-\sqrt3/2,3/2),\quad
  \bm\delta_z=(0,1),\\
  &\bm r_A(x,y)=x\bm a_1+y\bm a_2,\quad
  \bm r_B(x,y)=\bm r_A(x,y)+\bm\delta_z.
 \end{aligned}
 \label{eq:app-honeycomb-embedding}
\end{equation}
Thus $A(x,y)B(x,y)$ is a vertical $z$ bond with $A$ below $B$.
The $x$ and $y$ bonds from $A(x,y)$ end at $B(x-1,y)$ and
$B(x,y-1)$, respectively.  Taking the boundary of hexagon $p(x,y)$ in
clockwise order, the plaquette convention is
\begin{align}
 W_{p(x,y)}={}&
 \sigma^x_{A(x,y)}\sigma^y_{B(x,y)}
 \sigma^z_{A(x+1,y)}\nonumber\\[-2pt]
 &\times\sigma^x_{B(x+1,y-1)}
 \sigma^y_{A(x+1,y-1)}\sigma^z_{B(x,y-1)}.
 \label{eq:app-plaquette-coordinate-convention}
\end{align}
All coordinates below are understood modulo
$\bm T_1=(3,0)$ and $\bm T_2=(0,3)$.  We choose the crystal translation
$w_{p(x,y)}=+1$ if and only if $x-y\equiv0\pmod3$, with $w_p=-1$
on the other plaquettes [see Fig.~\ref{fig:app-dual-loops}(a)].  
For an elementary Pauli operator
$O$, let $\mathcal F(O)=\{p:\{O,W_p\}=0\}$ be its two adjacent plaquettes
whose fluxes it flips.  Since the lattice is trivalent, every site lies on three plaquettes, and
Eq.~\eqref{eq:app-plaquette-coordinate-convention} assigns it one bond color in
each.  A factor $\sigma^\beta_i$ of $W_p$ commutes with $\sigma^\alpha_i$ when
$\beta=\alpha$ and anticommutes otherwise, while the remaining five factors of
$W_p$ have no support on $i$; hence $\sigma^\alpha_i$ flips exactly the two
plaquettes whose assignment differs from $\alpha$, which is why $\mathcal F(O)$
always contains two elements.  
The assignments are
\begin{equation}
 \begin{aligned}
 A(x,y):&\ \ x\ \text{in}\ p(x,y),\ \
 z\ \text{in}\ p(x{-}1,y),\\
 &\ \ y\ \text{in}\ p(x{-}1,y{+}1),\\
 B(x,y):&\ \ y\ \text{in}\ p(x,y),\ \
 x\ \text{in}\ p(x{-}1,y{+}1),\\
 &\ \ z\ \text{in}\ p(x,y{+}1),
 \end{aligned}
 \label{eq:app-pauli-dual-incidence}
\end{equation}
so that, for example,
$\mathcal F(\sigma^y_{A(0,0)})=\{p(0,0),p(-1,0)\}$, the two plaquettes whose
assignment differs from $y$, namely $x$ and $z$.
In these coordinates,
\begin{equation}
 \begin{aligned}
 \mathcal T_1={}&
 \sigma^y_{A(0,0)}\sigma^x_{A(0,-1)}
 \sigma^z_{B(0,1)}\sigma^x_{B(0,0)},\\
 \mathcal T_2={}&
 \sigma^y_{A(0,0)}\sigma^z_{A(-1,0)}
 \sigma^y_{B(1,0)}\sigma^x_{B(0,0)}.
 \label{eq:app-dual-loops}
 \end{aligned}
\end{equation}
Substituting the successive factors of each string into
Eq.~\eqref{eq:app-pauli-dual-incidence} and following the dual path on the
cover, starting at $(0,0)$, gives
\begin{align}
 \widetilde{\mathcal C}_1:\quad
 &(0,0)\to(-1,0)\to(-1,-1)\to(0,-2)\to(0,-3),
 \nonumber\\[-2pt]
 &\hspace{30pt}\Delta\widetilde{\mathcal C}_1=(0,-3)=-\bm T_2,
 \label{eq:app-dual-lift-t1}\\
 \widetilde{\mathcal C}_2:\quad
 &(0,0)\to(-1,0)\to(-2,1)\to(-3,1)\to(-3,0),
 \nonumber\\[-2pt]
 &\hspace{30pt}\Delta\widetilde{\mathcal C}_2=(-3,0)=-\bm T_1.
 \label{eq:app-dual-lift-t2}
\end{align}
Each factor generates one step between hexagon centers on the dual
triangular lattice: the first creates a flux-defect pair, the intermediate
factors transport one endpoint, and the last annihilates the pair.  Every
plaquette is therefore flipped an even number of times, so both strings commute
with every local $W_p$ and preserve the crystal flux sector.  The nonzero
displacements in Eqs.~\eqref{eq:app-dual-lift-t1} and
\eqref{eq:app-dual-lift-t2} show that the strings are closed but
noncontractible on the quotient torus [see Fig.~\ref{fig:app-dual-loops}].   
Their intersections with the two direct cycles give the exact Pauli algebra
\begin{equation}
 \begin{aligned}
 \{\mathcal T_1,\mathcal W_1\}={}&\{\mathcal T_2,\mathcal W_2\}=0,\\
 [\mathcal T_1,\mathcal W_2]={}&[\mathcal T_2,\mathcal W_1]
 =[\mathcal T_1,\mathcal T_2]=0,
 \label{eq:app-dual-loop-algebra}
 \end{aligned}
\end{equation}
where $\mathcal W_1,\mathcal W_2$ are the noncontractible Wilson loops with
eigenvalues $w_1,w_2=\pm1$.  Thus $\mathcal T_\mu$ flips $w_\mu$ and preserves
the other.  These relations alone do not exclude the string creating a matter excitation in
the target Wilson sector.
 
To establish ground-space invariance, let
\begin{equation}
 \Pi_{\rm cr}=\prod_p\frac{1+w_p^{\rm cr}W_p}{2}
 \label{eq:app-crystal-projector}
\end{equation}
project onto the chosen local crystal-flux pattern.  Since
$[\mathcal T_\mu,W_p]=0$, one has
$[\mathcal T_\mu,\Pi_{\rm cr}]=0$.  Moreover,
$W_p\Pi_{\rm cr}=w_p^{\rm cr}\Pi_{\rm cr}$, and hence the crystal-sector
Hamiltonian,
\begin{align}
 H_{\rm cr}\equiv H_{\rm DW}^{\rm spin}\Pi_{\rm cr}
 &=-J_0\sum_e
 B_e\bigl(1-w_{p_e}^{\rm cr}w_{q_e}^{\rm cr}\bigr)\Pi_{\rm cr} \nonumber\\
 &=-2J_0\sum_{e\in E_{\rm act}}B_e\Pi_{\rm cr},
\end{align}
where $E_{\rm act}$ contains precisely the boundaries of the disjoint
crystal hexagons.  

The remaining commutators can be checked exactly in the Pauli algebra.
One verifies $[\mathcal T_\mu,B_e]=0$ factor by factor for every active
bond in Figs.~\ref{fig:app-dual-loops}(b)\&(c), and hence, 
\begin{equation}
 [\mathcal T_\mu,H_{\rm DW}^{\rm spin}]\Pi_{\rm cr}=0,
 \qquad
 [\mathcal T_\mu,H_{\rm cr}]=0.
 \label{eq:app-dual-loop-Hcomm}
\end{equation}
Note that this identity is restricted to the selected crystal sector.
Let $|\Psi_{w_1w_2}\rangle$ be the normalized crystal ground state in a
fixed Wilson sector.  Equation~\eqref{eq:app-dual-loop-Hcomm} gives
\begin{equation}
 H_{\rm DW}^{\rm spin}\mathcal T_\mu|\Psi_{w_1w_2}\rangle
 =E_0\mathcal T_\mu|\Psi_{w_1w_2}\rangle.
 \label{eq:app-dual-loop-energy}
\end{equation}
Combining this equality with Eq.~\eqref{eq:app-dual-loop-algebra}, and using
the uniqueness of the physical Majorana vacuum in each crystal-flux and
Wilson sector at $J_0<0$, yields
\begin{equation}
 \begin{aligned}
 \mathcal T_1|\Psi_{w_1w_2}\rangle
 &=e^{\ii\phi_1(w_1,w_2)}|\Psi_{-w_1,w_2}\rangle,\\
 \mathcal T_2|\Psi_{w_1w_2}\rangle
 &=e^{\ii\phi_2(w_1,w_2)}|\Psi_{w_1,-w_2}\rangle.
 \label{eq:app-dual-loop-ground-mapping}
 \end{aligned}
\end{equation}
The Pauli strings are unitary, so the states on the left-hand side are
nonzero and normalized.  Hence
\begin{equation}
 [\mathcal T_\mu,P_{\mathcal G}]=0,
 \qquad
 (1-P_{\mathcal G})\mathcal T_\mu P_{\mathcal G}=0,
 \label{eq:app-dual-loop-ground-invariance}
\end{equation}
which is the required statement that the dual strings act within the
ground space.
 
Because $\mathcal T_\mu^2=1$ and
$[\mathcal T_1,\mathcal T_2]=0$, the phases in
Eq.~\eqref{eq:app-dual-loop-ground-mapping} can be absorbed into the four
sector states.  In the resulting basis, the projected operators act as
\begin{equation}
 \mathcal W_1={}Z\otimes \mathbb I,
 \quad\mathcal W_2={}\mathbb I \otimes Z,
 \quad\mathcal T_1={}X\otimes \mathbb I,
 \quad\mathcal T_2={}\mathbb I \otimes X,
 \label{eq:app-projected-loop-algebra}
\end{equation}
where all four operators are projected onto $\mathcal G$.  The maximum ED
residual over unitarity, involution, and the relations above is
$1.6\times10^{-15}$.  The four states therefore realize two conjugate logical
Pauli pairs: the dual strings connect rather than excite them.
Equation~\eqref{eq:app-projected-loop-algebra} is itself size independent: four
involutions obeying Eq.~\eqref{eq:app-dual-loop-algebra} generate the two-qubit
Pauli group, whose four-dimensional irreducible representation is unique up to
unitary equivalence.  The $N_p=9$ calculation supplies the remaining input in
the original spin variables---explicit strings, the uniqueness of the physical
Majorana vacuum at $J_0<0$, and the commutators of
Eqs.~\eqref{eq:app-dual-loop-algebra} and~\eqref{eq:app-dual-loop-Hcomm}---with
no reference to the mapping of Sec.~\ref{sec:toriccode}.  It is therefore an
independent check of that mapping rather than a corollary of it, and the
mapping in turn supplies the logical operators at every commensurate size.
Each factor is a single-site Pauli, so its commutators are fixed by its color
and local environment alone; since a string contains only the three factor
types above, three verifications suffice, and a larger torus adds only further
transport factors.  Note that $N_p=9$ is the only commensurate size at which
this check is feasible in the spin representation, the next square commensurate
torus $N_p=36$ carrying $N_s=72$ spins.  
These results establish the logical ground-space structure at the loop point and do not,
by themselves, establish stability under generic perturbations.

\bibliography{reference.bib}

\end{document}